\documentclass[prd,aps,showpacs,preprintnumbers,amsmath,amssymb,nofootinbib]{revtex4}
\usepackage{subeqnarray}
\usepackage{graphicx}
\usepackage{color}
\usepackage{caption}
\usepackage{multirow}
\usepackage{hhline}
\usepackage{slashed}
\usepackage{bbm}
\usepackage{xspace}
\usepackage{float}
\usepackage{epstopdf}
\usepackage{textcomp}
\usepackage{anysize}
\usepackage{bold-extra}
\usepackage{enumerate}
\usepackage{lscape}
\usepackage{mathrsfs}
\usepackage[final]{pdfpages}
\usepackage{rotating}
\usepackage{subfig}
\usepackage{url}
\usepackage{wrapfig}
\usepackage{fancyhdr}
\usepackage{verbatim}
\usepackage[export]{adjustbox}
\usepackage{hyperref}
\allowdisplaybreaks[4]

\begin{document}

\newcommand{\rem}[1]{{\bf [#1]}}
\newcommand{\gsim}{ \mathop{}_ {\textstyle \sim}^{\textstyle >} }
\newcommand{\lsim}{ \mathop{}_ {\textstyle \sim}^{\textstyle <} }
\newcommand{\vev}[1]{ \left\langle {#1}  \right\rangle }
\newcommand{\bear}{\begin{array}}
\newcommand {\eear}{\end{array}}

\newcommand{\beq}{\begin {equation}}
\newcommand{\eeq}{\end   {equation}}
\newcommand{\bea}{\begin {eqnarray}}
\newcommand{\eea}{\end   {eqnarray}}
\newcommand{\baa}{\begin {array}   }
\newcommand{\eaa}{\end   {array}   }
\newcommand{\bit}{\begin {itemize} }
\newcommand{\eit}{\end   {itemize} }
\newcommand{\be }{\begin {equation}}
\newcommand{\ee }{\end   {equation}}
\newcommand{\nn }{\nonumber        }

\newcommand{\bef}{\begin{figure}}
\newcommand {\eef}{\end{figure}}
\newcommand{\bec}{\begin{center}}
\newcommand {\eec}{\end{center}}
\newcommand{\non}{\nonumber}
\newcommand {\eqn}[1]{\beq {#1}\eeq}
\newcommand{\la}{\left\langle}
\newcommand{\ra}{\right\rangle}
\newcommand{\ds}{\displaystyle}
\newcommand{\red}{\textcolor{red}}

\newcommand{\met}{\not{\!\!{\rm E}}_{T}}
\newcommand{\AFB}{A_{FB}}
\newcommand{\eeha}{e^{+}e^{-}\rightarrow h \gamma}
\newcommand{\hlla}{h\rightarrow \gamma\gamma,Z\gamma \rightarrow l^{+}l^{-}\gamma}
\newcommand{\eebbah}{e^{+}e^{-}\rightarrow h \gamma \rightarrow b \bar{b}\gamma}
\newcommand{\eebbaz}{e^{+}e^{-}\rightarrow \gamma^{*}\gamma,Z^{*}\ \gamma \rightarrow b \bar{b}\gamma}
\newcommand{\GeV}{~\mathrm{GeV}}
\global\long\def\fl{\mathcal{F}_{L}}

\newcommand{\slx}{s_L}

\def\SEC#1{Sec.~\ref{#1}}
\def\FIG#1{Fig.~\ref{#1}}
\def\EQ#1{Eq.~(\ref{#1})}
\def\EQS#1{Eqs.~(\ref{#1})}
\def\lrf#1#2{ \left(\frac{#1}{#2}\right)}
\def\lrfp#1#2#3{ \left(\frac{#1}{#2} \right)^{#3}}
\def\GEV#1{10^{#1}{\rm\,GeV}}
\def\MEV#1{10^{#1}{\rm\,MeV}}
\def\KEV#1{10^{#1}{\rm\,keV}}
\def\REF#1{Ref.~\cite{#1}}
\def\APPEN#1{Appendix~\ref{#1}}
\def\lrf#1#2{ \left(\frac{#1}{#2}\right)}
\def\lrfp#1#2#3{ \left(\frac{#1}{#2} \right)^{#3}}
\def\OG#1{ {\cal O}(#1){\rm\,GeV}}



\title{Probing CP-violating $h\bar tt$ coupling in $e^{+}e^{-}\rightarrow h \gamma$}
\author{Gang Li$^{1,}$\footnote{Email: gangli@pku.edu.cn}, Hao-Ran Wang$^{1,}$\footnote{Email: haorwang@pku.edu.cn}, Shou-hua Zhu$^{1,2,3,}$\footnote{Email: shzhu@pku.edu.cn}}

\affiliation{$^1$Department of Physics and State Key Laboratory of Nuclear Physics
and Technology, Peking University, Beijing 100871, China\\
$^2$Collaborative Innovation Center of Quantum Matter, Beijing, China\\
$^3$Center for High Energy Physics, Peking University, Beijing 100871, China
}
\date{\today}
\begin{abstract}
\vspace*{0.5cm}
We investigate the possibility of probing the CP-violating $h\bar tt$ coupling in the process $\eeha$ at the future high luminosity $e^{+}e^{-}$ colliders.
Our numerical results show that the cross section for this process can be significantly increased for the allowed CP phase $\xi$ and center of mass energy. 
For example the cross section is about 10 times of that in the standard model (SM) for $\sqrt{s}=350\GeV$ and $\xi =3\pi/5$ (see text for $\xi$ definition).
The simulation for the signal process $\eebbah$ and its backgrounds shows that the signal significance can reach about $5\sigma$ and more than $2.1\sigma$ for 
$\sqrt{s}=350\GeV,\ 500\GeV$ respectively, with the integrated luminosity $\mathcal{L}=3~\text{ab}^{-1}$ and $\xi\in[\pi/2,3\pi/5]$. For $\mathcal{L}=10~\text{ab}^{-1}$,
the signal significance can be greater than $5\sigma$ for $\sqrt{s}=350\GeV$ and  about $4\sigma$ for $\sqrt{s}=500\GeV$ with the CP phase $\xi\in[\pi/2,3\pi/5]$. Besides the cross section enhancement, the CP-violating $h\bar tt$ coupling will induce a
forward-backward asymmetry $\AFB$ which is absent in the SM and is a clear signal of new CP violation. Compared with the $\AFB$ in the Higgs decay $h\rightarrow l^{+}l^{-}\gamma$, the $\AFB$ can be greatly enhanced in the production process. For example $\AFB$ can reach -0.55 for $\xi=\pi/4$ and $\sqrt{s}=500\GeV$. Due to the large backgrounds, the significance of the expected $\AFB$ can be only observed at $1.68\sigma$ with $\mathcal{L}=10~\text{ab}^{-1}$ and $\sqrt{s}=500\GeV$. It is essential to trigger the single photon in the final state to separate the bottom jets arising from scalar or vector bosons,
in order to isolate the signal from the backgrounds more efficiently.

\end{abstract}
\maketitle

\newpage

\section{Introduction}
\label{sec:1}
The observation of the 125 GeV Higgs boson at the Large Hadron Collider (LHC)~\cite{Aad:2012tfa,Chatrchyan:2012ufa} marked a milestone in particle physics.  Consequently detailed measurements of the discovered Higgs boson properties have become one of the main priorities of the LHC and future colliders, 
in order to verify whether it is the SM Higgs boson or not. In fact, there are various motivations for new physics beyond the SM (BSM)~\cite{Zhu:2014hda}. In the SM, charge conjugation-parity (CP) violation is described by a $3\times3$ quark mixing matrix, the well-known Cabibbo-Kobayashi-Maskawa (CKM) matrix~\cite{Cabibbo:1963yz,Kobayashi:1973fv}, with a single complex phase in the gauge sector, while the Higgs sector is CP-conserved. However, the CP violation in the SM can not account for the origin of the baryon asymmetry of the universe (BAU)~\cite{Gavela:1993ts, Huet:1994jb} which is characterized by the baryon-to-entropy ratio $n_{B}/s\simeq (8.59\pm0.10)\times 10^{-11}$~\cite{Ade:2013zuv, Agashe:2014kda}. Thus it is necessary to look for new sources of CP violation.

Phenomenologically there are already many works on studying the CP-violating Higgs couplings~\cite{Godbole:2007cn,Cao:2009ah, Belyaev:2015xwa, Delaunay:2013npa, Bishara:2013vya, Chen:2014ona, Korchin:2013ifa,Korchin:2014kha, Farina:2015dua, Gunion:1996vv, BhupalDev:2007is,Godbole:2011hw, Kolodziej:2013ima, Ellis:2013yxa, Khatibi:2014bsa, He:2014xla, Boudjema:2015nda, Nishiwaki:2013cma,Artoisenet:2013puc, Maltoni:2013sma, Demartin:2014fia, Kobakhidze:2014gqa, Demartin:2015uha, Harnik:2013aja, Belusca-Maito:2014dpa, Dwivedi:2015nta,Brod:2013cka, Mao:2014oya, Inoue:2014nva, Chen:2015gaa}. In general, the CP-odd gauge-Higgs couplings are generated from higher dimension operators~\cite{Buchmuller:1985jz} and usually smaller compared to the CP-even couplings. However $h\gamma\gamma,\ hZ\gamma$ couplings are not present at the tree-level, so CP-even and CP-odd components can be of the similar magnitude.
Especially the $h\gamma\gamma,hZ\gamma$ couplings can be induced by the CP-even and CP-odd Yukawa couplings~\cite{Korchin:2014kha}, which are not suppressed and comparably large~\cite{Inoue:2014nva, Chen:2015gaa}. Due to the interference of the fermion loops and $W$ boson loops, the $h\gamma\gamma,hZ\gamma$ couplings are sensitive to the Higgs coupling to $t$ quarks ~\cite{Shen:2015pha, Korchin:2014kha, Chen:2015rha}.

The forward-backward asymmetry ($\AFB$) as a consequence of parity violation has been studied extensively in $Z$ physics at the LEP~\cite{ALEPH:2005ab}. In~\cite{Chen:2014ona, Korchin:2014kha, Belyaev:2015xwa}, the authors proposed to measure the CP violation in Higgs decays using this observable. Since the intermediate $Z$ boson is dominantly on-shell but the photon is virtual and the imaginary part is proportional to the $\gamma-Z$ interference term, the $\AFB$ of the process $h\rightarrow l^{+}l^{-}\gamma$ is suppressed by a factor of $\Gamma_{Z}/m_{Z}\sim 3\%$~\cite{Chen:2014ona,Korchin:2014kha}. However, the $\AFB$ will be greatly enhanced in the production process $e^+e^- \rightarrow h \gamma$, where the $h\gamma\gamma$ and $hZ\gamma$ couplings can have large imaginary parts (due to resonance effects)~\cite{Gounaris:2015tna}. In this paper, we will discuss the possibility of probing the CP property of $h\bar{t}t$ coupling in $\eeha$ at the future $e^{+}e^{-}$ colliders.

The paper is organized as follows. In~\SEC{sec:formalism}, we will present the one-loop amplitude of $\eeha$ and the analytical formula of $\AFB$ as well as the CP symmetry of the helicity amplitude. In~\SEC{sec:numerical}, we give the numerical results with the help of FeynArts/FormCalc/LoopTools~\cite{Hahn:2000kx, Hahn:1998yk} for some benchmark center-of-mass (c.m.) energies and CP phases. In~\SEC{sec:collider}, the collider simulation of signal and backgrounds is presented. The analytical expressions of scalar functions, $W$ boson loop functions and box diagrams contributions and the derivation of the significance of $A_{FB}$ are collected in the appendices.

\section{Formalism}
\label{sec:formalism}
We first parameterize the CP-violating $h\bar{t}t$ coupling as~\cite{Ellis:2013yxa}
\beq
\label{eq:htt}
\mathcal{L}_{htt}=-\frac{m_{t}}{v}\kappa_{t}\bar{t}(\cos\xi_t+i\sin\xi_t\gamma_{5})t,
\eeq
where $v=246\GeV$ is the vacuum expectation value (vev) of the Higgs field, $\kappa_{t}\in\mathbb{R}$ describes the magnitude of $h\bar tt$ interaction and $\xi\equiv\xi_t\in (-\pi,\pi]$. The CP-even and CP-odd cases correspond to $\xi=0$ and $\xi=\pi/2$, respectively. In particular, the SM Higgs boson has $\kappa_{t}=1$ and $\xi=0$. The signal strengths measured in $gg\rightarrow h$ and $h\rightarrow \gamma\gamma$ have constrained the CP phase $|\xi|\leq 3\pi/5$ at $95\%$ C.L.~\cite{Ellis:2013yxa, Nishiwaki:2013cma}\footnote{ATLAS Collaboration has performed a global analysis of the Higgs couplings~\cite{Aad:2015tna}, which may give a stronger constraint on the $h\bar{t}t$ coupling, however it depends on the assumption of the $hgg$ couplings as well so the conclusions in our paper will not change. }. On the other hand, the electric dipole moment (EDM) of the electron $d_{e}$ constrains the upper limit on the CP-odd coupling $|\kappa_{t}\sin\xi|<0.01$ with the assumption of SM-like $h\bar ee$ coupling~\cite{Brod:2013cka}. However, this can be evaded in various new physics (NP) models~\cite{Inoue:2014nva, Cheung:2014oaa, Bian:2014zka}.  We will suppose $\kappa_{t}=1$ and $\xi\in[0,3\pi/5]$ in this paper.

\begin{figure}
  \centering
   \includegraphics[width=0.7\textwidth]{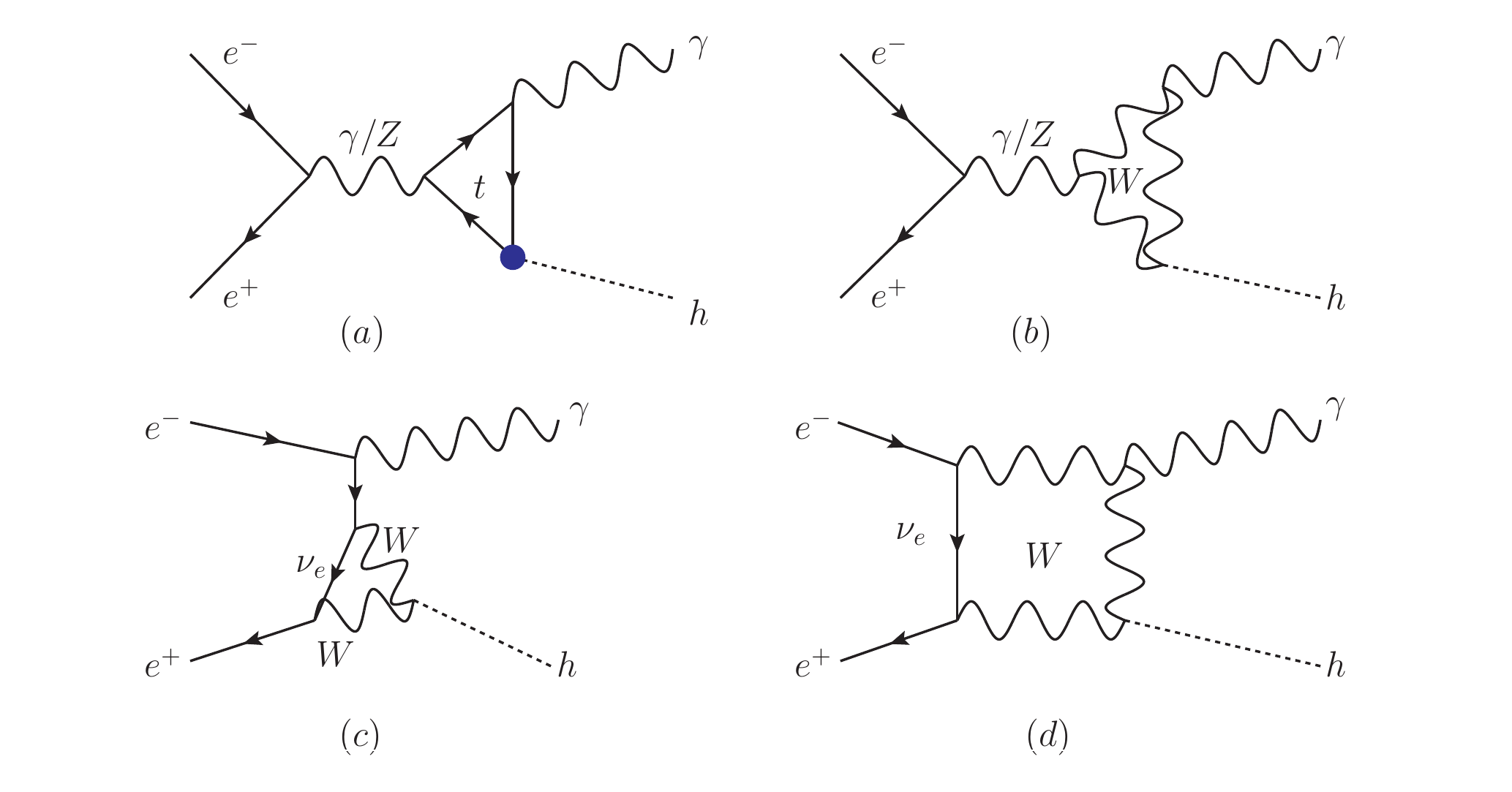}
   \caption{Representative diagrams for $\eeha$. $(a)(b)$ denote the s-channel vertex diagrams, $(c)(d)$ denote t-channel vertex diagrams and $W/$ neutrino box diagrams, respectively. The CP-violating $h\bar tt$ coupling is marked with the blue dot.}
 \label{fig:feynmandiagram}
   \end{figure}

The process $\eeha$ has been investigated in the SM~\cite{Bergstrom:1985hp, Barroso:1985et, Dicus:1995rc} and beyond the SM~\cite{ Djouadi:1996ws,Hagiwara:1993sw, GonzalezGarcia:1999fq, Gounaris:2015tna, Cao:2015fra,Cao:2015iua,Ren:2015uka, Biswas:2015sha}. The tree-level amplitude is suppressed by the small $h\bar{e}e$ Yukawa coupling and can be safely ignored~\cite{Barroso:1985et}.~\FIG{fig:feynmandiagram} shows the representative diagrams for $\eeha$ at one-loop level\footnote{The $Z/$electron box diagram is obtained by replacing $W$ boson and neutrino with $Z$ boson and electron, and attaching the photon to electron line since there is no $\gamma ZZ$ interaction at tree-level.}, where only the $h\bar tt$ Yukawa coupling is assumed to be CP-violated and contained in $h\gamma\gamma$ and $hZ\gamma$ vertex diagrams and $\gamma/Z-h$ mixing diagrams. The latter is the higher order correction to $h\bar ee$ coupling, which is also neglected. The $h\gamma\gamma$ and $hZ\gamma$ effective couplings take the form as~\cite{Djouadi:1996ws}
\beq
\Gamma_{V}^{\mu\nu}=G_{1}^{V}g^{\mu\nu}+G_{3}^{V}k^{\nu}q^{\mu}+G_{6}^{V}\epsilon^{\mu\nu\alpha\beta}q_{\alpha}k_{\beta},
\eeq
where $q$ is the four-momentum of the intermediate propagator $V=\gamma,Z$ and the coefficients $G_{1,3,6}^{\gamma/Z}$ are generally complex,
 \bea
 \label{eq:G-F}
 G_{i}^{\gamma}&=&\chi\frac{e^3m_{W}}{s_{W}}[F_{i}^{\gamma,W}-\sum_{f}4Q_{f}^{2}N_{c}\frac{m_{f}^{2}}{m_{W}^{2}}F_{i}^{f}],\nn\\
  G_{i}^{Z}&=&-\chi\frac{e^3m_{W}}{c_{W}s_{W}^{2}}[F_{i}^{Z,W}+\sum_{f}2Q_{f}N_{c}\frac{m_{f}^{2}}{m_{W}^{2}}g_{V}^{f}F_{i}^{f}],
 \eea
where $\chi=1,6$ for $i=1$ while $\chi=-1$ for $i=3$. $Q_{f}$, $m_{f}$ and $N_{c}$ are the electric charge, mass of the fermion and color factor, respectively, and the vector part of $Zf\bar{f}$ coupling $g_{V}^{f}=T_{3}^{f}-2\sin^{2}\theta_{W}Q_{f}$ with the third component weak isospin $T_{3}^{f}=\pm\frac{1}{2}$ for the left-handed fermion $f$. The Higgs boson is C-even and the photon is C-odd, only the C-odd (vector) part of the $Zf\bar{f}$ coupling $g_{V}^{f}$ contributes~\cite{Bergstrom:1985hp, Barroso:1985et} even though the $h\bar{t}t$ coupling is CP-violated~\cite{Korchin:2014kha}. The fermion-loop contribution functions $ F_{1}^{f}$, $F_{3}^{f}$ and $F_{6}^{f}$ are given by ($\xi_{f}=0$ for $f\neq t$)
 \bea
 \label{eq:fermionformfactor}
  F_{1}^{f}&=&\frac{1}{2(s-m_{h}^{2})}\cos\xi_{f}[(m_{h}^{2}-s)(-4m_{f}^{2}+m_{h}^{2}-s)C_{0}+2s(B_{0}(s)-B_{0}(m_{h}^{2}))+2(s-m_{h}^{2})],\nn\\
  F_{3}^{f}&=&\frac{1}{(m_{h}^{2}-s)^{2}}\cos\xi_{f}[(m_{h}^{2}-s)(-4m_{f}^{2}+m_{h}^{2}-s)C_{0}+2s(B_{0}(s)-B_{0}(m_{h}^{2}))+2(s-m_{h}^{2})],\nn\\
  F_{6}^{f}&=&-\sin\xi_{f}C_{0},
 \eea
satisfying the gauge invariance $k_{\mu}\cdot(F_{1}^{f}g^{\mu\nu}-F_{3}^{f}k^{\nu}q^{\mu})=F_{3}^{f}k_{\mu}\cdot(k\cdot q g^{\mu\nu}-k^{\nu}q^{\mu})=0$ with $k\cdot q=(s-m_h^2)/2$ and $\epsilon^{\mu\nu\alpha\beta}k_{\mu}k_{\beta}q_{\alpha}=0$. In \EQ{eq:fermionformfactor}, $C_{0}\equiv C_{0}(0,s,m_{h}^{2},m_{f}^{2},m_{f}^{2},m_{f}^{2}),B_{0}(s)\equiv B_{0}(s,m_{f}^{2},m_{f}^{2}),B_{0}(m_{h}^{2})\equiv B_{0}(m_{h}^{2},m_{f}^{2},m_{f}^{2})$ with the analytical expressions being given in Appendix~\ref{Appdendix:analyticalexpression}.

The $W$-loop contribution functions $ F_{i}^{\gamma,W}$ and $F_{i}^{Z,W}(i=1,3)$ in the Feynman gauge can be found in~\cite{Barroso:1985et,Djouadi:1996ws} and are collected in Appendix~\ref{loopfunctions}, while $ F_{6}^{\gamma,W}=F_{6}^{Z,W}=0$. As first pointed out in \cite{Barroso:1985et} the $W$-loop contribution to the $hZ\gamma$ vertex diagram contains divergence which is cancelled by the divergent part of the $W$-loop contributions to the $Z-\gamma$ mixing, while fermion-loop contributions to the $Z-\gamma$ mixing are proportional to the final photon momentum squared and therefore vanishes for photon being on-shell.

Following with the notations in~\REF{Djouadi:1996ws}, the amplitude $\mathcal{M}$ can be decomposed as
\beq
\mathcal{M}=\frac{1}{16\pi^2}\sum_{i=1,2,3,6}\sum_{a=\pm}\Lambda_{i}^{a}C_{i}^{a},
\eeq
and the form factor $C_{i}^{\pm}$ sum all the diagrams
\beq
\label{eq:formfactors}
C_{i}^{\pm}=C_{i}^{\gamma\pm}+C_{i}^{Z\pm}+C_{i}^{e\pm}+C_{i}^{\mathrm{box}\pm},
\eeq
where the form factors $C_{i}^{\gamma\pm},C_{i}^{Z\pm},C_{i}^{e\pm},C_{i}^{\mathrm{box}\pm}$ denote the contributions of the $h\gamma\gamma$ vertex diagrams, the $hZ\gamma$ vertex diagrams, the t-channel $h\bar{e}e$ vertex corrections and the box diagrams, respectively. Contributions from $Z-\gamma$ mixing diagrams are also included in $C_{i}^{Z\pm}$. The matrix elements are given by
\bea
\label{eq:matrixelement}
\Lambda_{1}^{\pm}&=&\bar{v}(p_{+})(1\pm\gamma_{5})\gamma_{\nu}(g^{\mu\nu}k\cdot p_{-}-k^{\nu}p_{-}^{\mu})u(p_{-})\epsilon_{\mu}^{*}(k),\nn\\
\Lambda_{2}^{\pm}&=&\bar{v}(p_{+})(1\pm\gamma_{5})\gamma_{\nu}(g^{\mu\nu}k\cdot p_{+}-k^{\nu}p_{+}^{\mu})u(p_{-})\epsilon_{\mu}^{*}(k),\nn\\
\Lambda_{3}^{\pm}&=&\bar{v}(p_{+})(1\pm\gamma_{5})\gamma_{\nu}g^{\mu\nu}u(p_{-})\epsilon_{\mu}^{*}(k),\nn\\
\Lambda_{6}^{\pm}&=&\bar{v}(p_{+})(1\pm\gamma_{5})\gamma_{\nu}\epsilon^{\mu\nu\alpha\beta}q_{\alpha}k_{\beta}u(p_{-})\epsilon_{\mu}^{*}(k),
\eea
where $p_{\pm}$ and $k$ are the four-momenta of $e^{\pm}$ and the final photon, $\epsilon_{\mu}^{*}(k)$ denotes the polarization vector of the final photon and the Levi-Civita tensor $\epsilon^{0123}=1$. The correspondences between $G_{1,3,6}^{\gamma/Z}$ and $C_{1,2,3,6}^{\gamma/Z\pm}$ are
\bea
\label{eq:C-G}
C_{1}^{\gamma\pm}&=&C_{2}^{\gamma\pm}=\frac{e}{2s}G_{3}^{\gamma},\nn\\
C_{3}^{\gamma\pm}&=&-\frac{e}{2s}(G_{1}^{\gamma}+\frac{s-m_{h}^{2}}{2}G_{3}^{\gamma}),\nn\\
C_{6}^{\gamma\pm}&=&-\frac{e}{2s}G_{6}^{\gamma},\nn\\
C_{1}^{Z\pm}&=&C_{2}^{Z\pm}=-\frac{ez^{\pm}}{4s_{W}c_{W}(s-m_{Z}^2+im_{Z}\Gamma_{Z})}G_{3}^{Z},\nn\\
C_{3}^{Z\pm}&=&\frac{ez^{\pm}}{4s_{W}c_{W}(s-m_{Z}^2+im_{Z}\Gamma_{Z})}(G_{1}^{Z}+\frac{s-m_{h}^2}{2}G_{3}^{Z}),\nn\\
C_{6}^{Z\pm}&=&\frac{ez^{\pm}}{4s_{W}c_{W}(s-m_{Z}^2+im_{Z}\Gamma_{Z})}G_{6}^{Z},
\eea
where $s=q^2,z^{+}=g_{V}^{e}+g_{A}^{e},z^{-}=g_{V}^{e}-g_{A}^{e}$ with $g_{V}^{e}=-\frac{1}{2}+2\sin^{2}\theta_{W}$,\ $g_{A}^{e}=-\frac{1}{2}$, and $m_{Z}$, $\Gamma_{Z}$ are the mass and width of $Z$ boson, respectively. The t-channel and box diagram contributions are displayed in Appendix \ref{loopfunctions}. The gauge invariance of the $\eeha$ amplitude is maintained after summing over $W$-loop contributions to $C_{3}^{\gamma\pm},\ C_{3}^{Z\pm}$ and $C_{3}^{e\pm},\ C_{3}^{\mathrm{box}\pm}$~\cite{Djouadi:1996ws} (the fermion-loop contributions to $C_{3}^{\gamma\pm}$, $C_{3}^{Z\pm}$ are zero), namely the relations in \EQ{gi1}. 

The unpolarized differential cross section of the process $\eeha$ is
\beq
\label{eq:differentialcrosssection}
\frac{d\sigma}{d\cos\theta}=\frac{s-m_{h}^{2}}{64\pi s}\frac{1}{(16\pi^2)^2}\{u^2(|C_{1}^{+}|^2+|C_{1}^{-}|^2)+t^2(|C_{2}^{+}|^2+|C_{2}^{-}|^2)+ \mathcal{CPV}\},
\eeq
where $C_{i}^{\pm}$ are defined in~\EQ{eq:formfactors} and
\beq
\label{cpv1}
\mathcal{CPV}=-\frac{(s-m_{h}^2)^2}{2}[(\mathcal{F}_{1-}-\mathcal{F}_{1+})(1+\cos\theta)^2
-(\mathcal{F}_{2-}-\mathcal{F}_{2+})(1-\cos\theta)^2-(|C_{6}^{-}|^2+|C_{6}^{+}|^2)(1+\cos^2\theta)]
\eeq
with
\beq
\label{cpv2}
\mathcal{F}_{1\pm}=\frac{1}{2i}(C_{6}^{\pm*}C_{1}^{\pm}-C_{6}^{\pm}C_{1}^{\pm*}),\ \mathcal{F}_{2\pm}=\frac{1}{2i}(C_{6}^{\pm*}C_{2}^{\pm}-C_{6}^{\pm}C_{2}^{\pm*}).
\eeq
The kinematic variables $s,t$ and $u$ are defined as $s=(p_{+}+p_{-})^2$, $u=(p_{-}-k)^2=-\frac{1}{2}(s-m_{h}^2)(1+\cos\theta)$, $t=(p_{+}-k)^2=-\frac{1}{2}(s-m_{h}^2)(1-\cos\theta)$ and $\theta$ is the scattering angle between the directions of final photon and initial positron in the c.m. frame. The combination of the first two terms in \EQ{eq:differentialcrosssection} is symmetric in $\cos\theta$ even in the presence of CP-violating $h\bar{t}t$ coupling. While the last term, i.e. $\mathcal{CPV}$, can induce a forward-backward asymmetry which can be expressed as
\beq
\label{afbexact}
A_{FB}=\frac{\sigma_{F}-\sigma_{B}}{\sigma_{F}+\sigma_{B}}
\eeq
and
\beq
\sigma_{F}=\int_{0}^{1}\frac{d\sigma}{d\cos\theta}d\cos\theta,\
\sigma_{B}=\int_{-1}^{0}\frac{d\sigma}{d\cos\theta}d\cos\theta,
\eeq

\begin{figure}[!htb]
  \centering
 \includegraphics[width=0.4\textwidth]{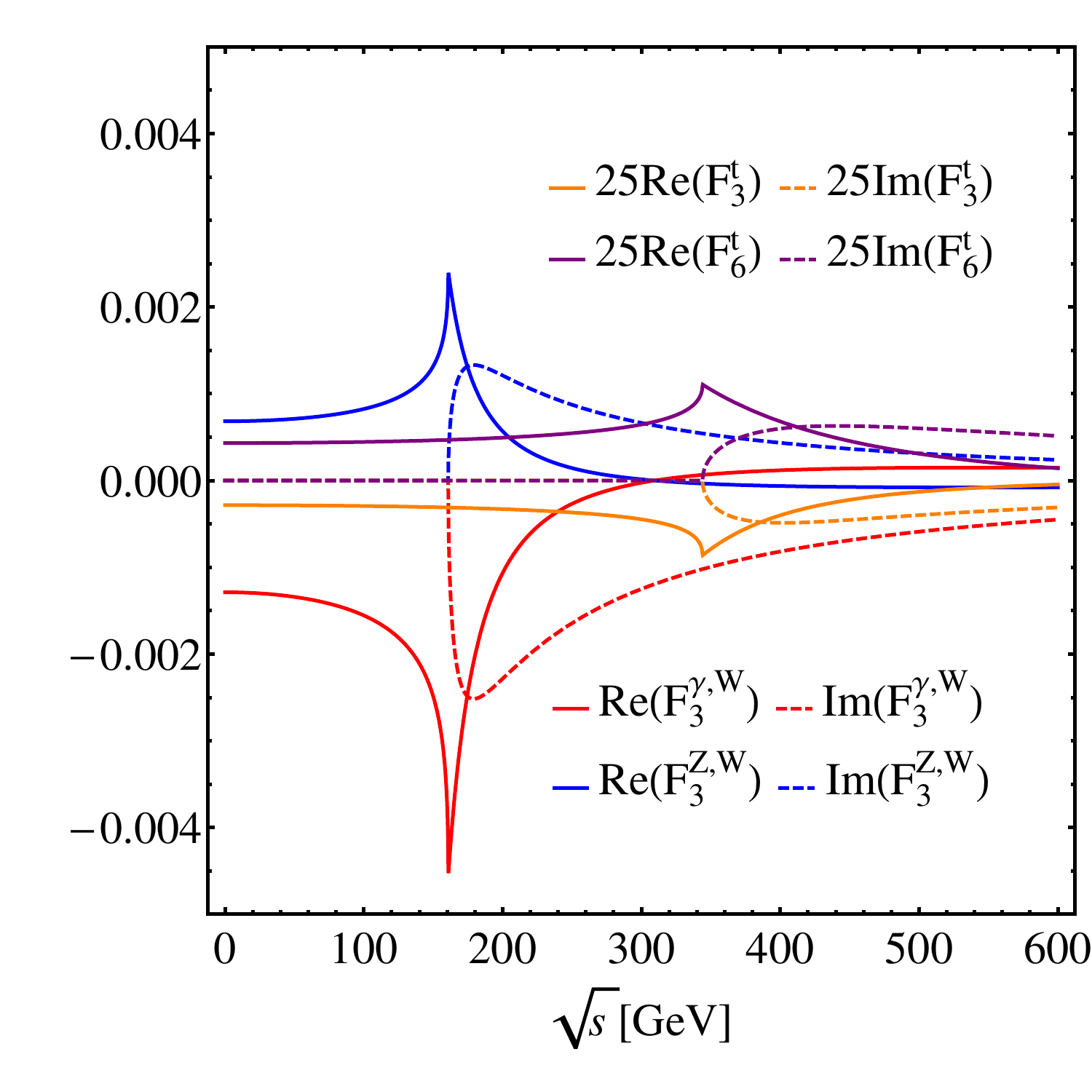}
   \caption{The energy dependences of various loop functions. For illustration, we have multiplied $F_{3}^{t}$ and $F_{6}^{t}$ with a factor of $4Q_t^2N_cm_t^2/m_W^2\simeq 25$ and chosen $\xi=0,\pi/2$, respectively.}.
     \label{fig:loopfunctions}
   \end{figure}

From the expressions in \EQS{eq:C-G}~\eqref{cpv1}~\eqref{cpv2}, the s-channel vertex diagram contribution to $\mathcal{CPV}$ is like $a+b\cos\theta+c\cos^2\theta$, where $a,b,c$ are generic coefficients.  Furthermore, $b$ is proportional to the difference of right- and left-handed couplings, i.e. the axial vector coupling $g_A^e$ of the $Z$ boson to $e^+e^-$. For nonzero $\mathcal{F}_{1\pm}$ and $\mathcal{F}_{2\pm}$, both the presence of CP phase $\xi$ and strong phase are required, the latter of which arises from the threshold effects with on-shell intermediate $W$ boson, $Z$ boson ($Z/$electron box diagram) and/or fermions in the loops. In \FIG{fig:loopfunctions}, we show the energy dependences of loop functions of the s-channel diagrams.
    \begin{figure}[!htb]
     \centering
 \includegraphics[width=0.3\textwidth]{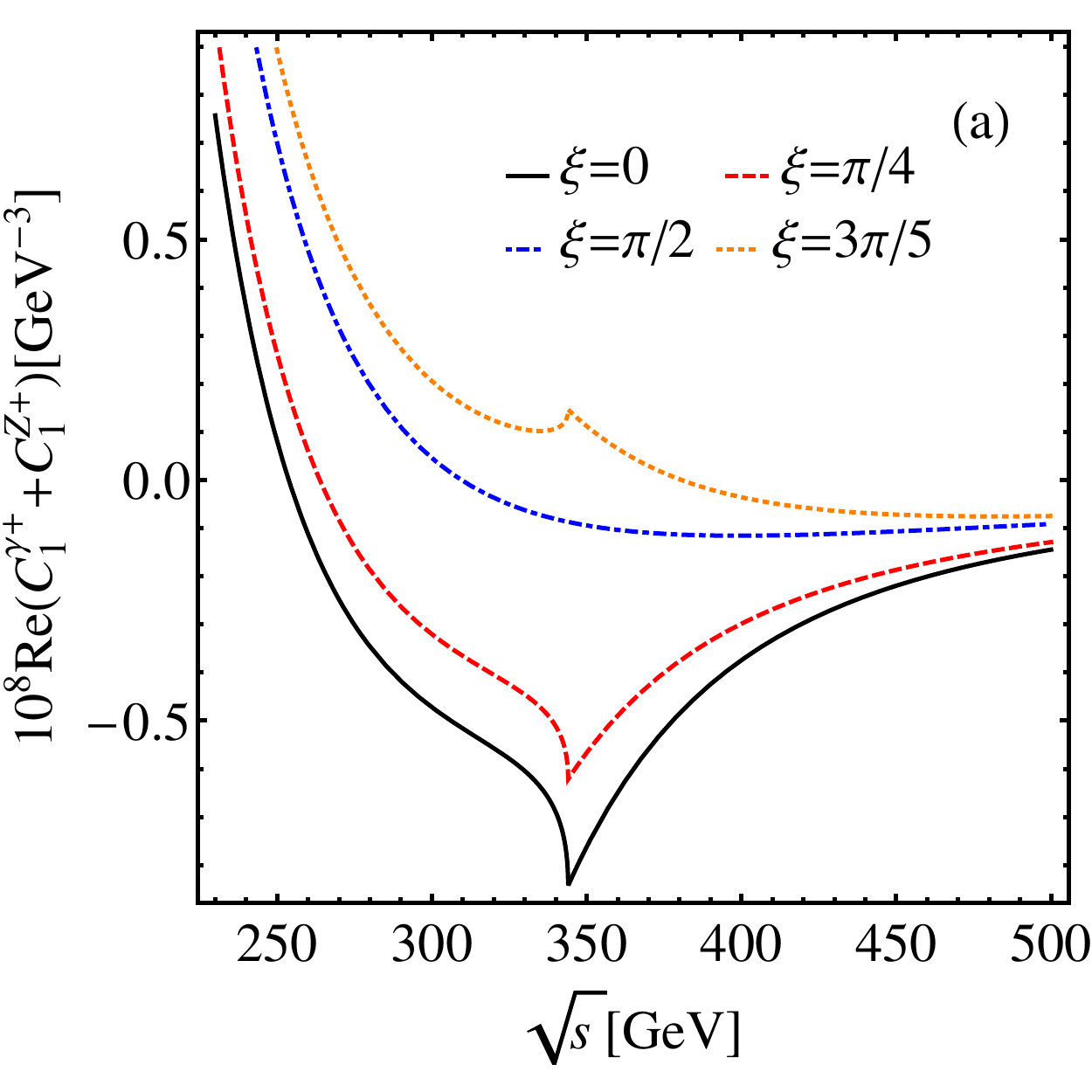}
 \includegraphics[width=0.3\textwidth]{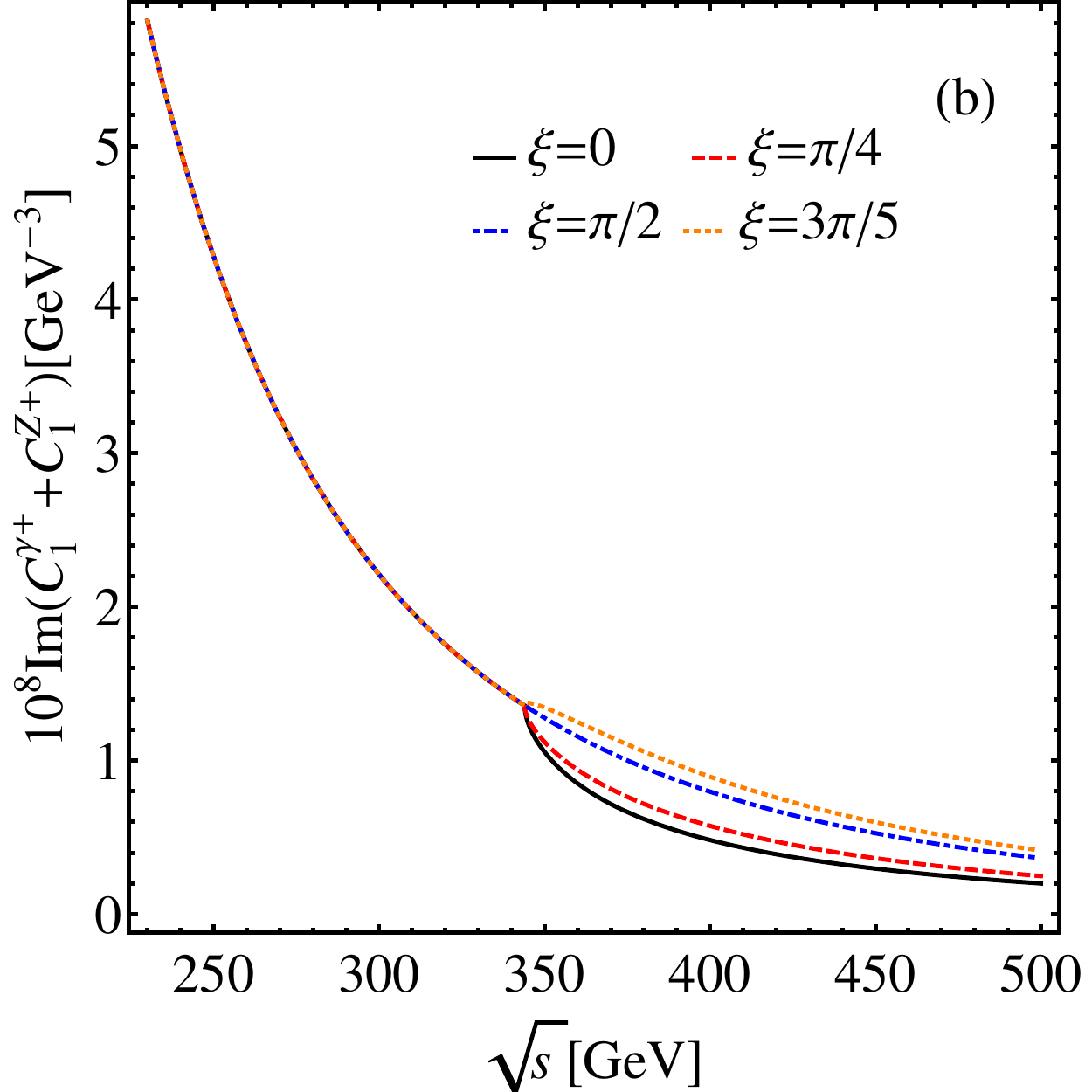}\\
  \includegraphics[width=0.3\textwidth]{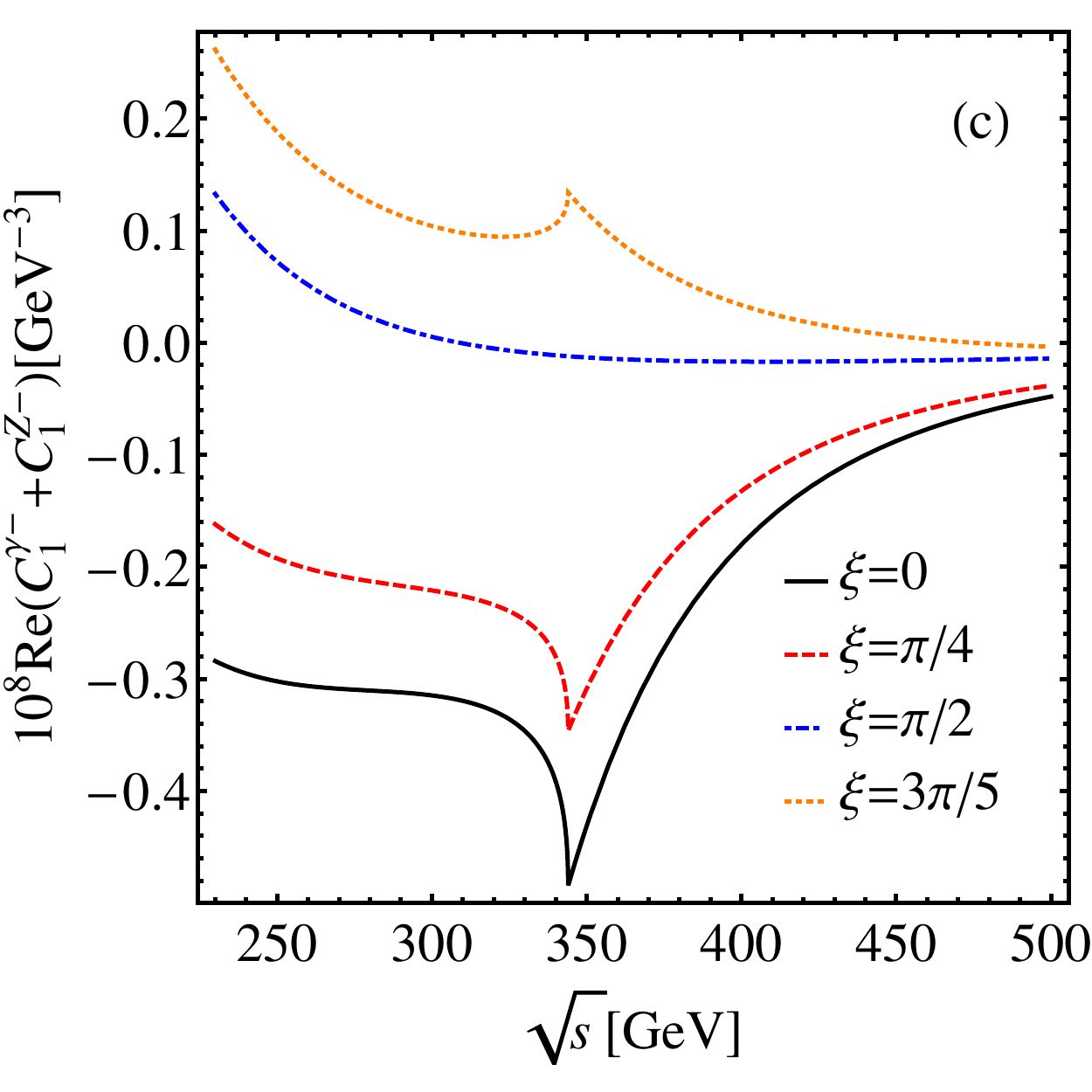}
   \includegraphics[width=0.3\textwidth]{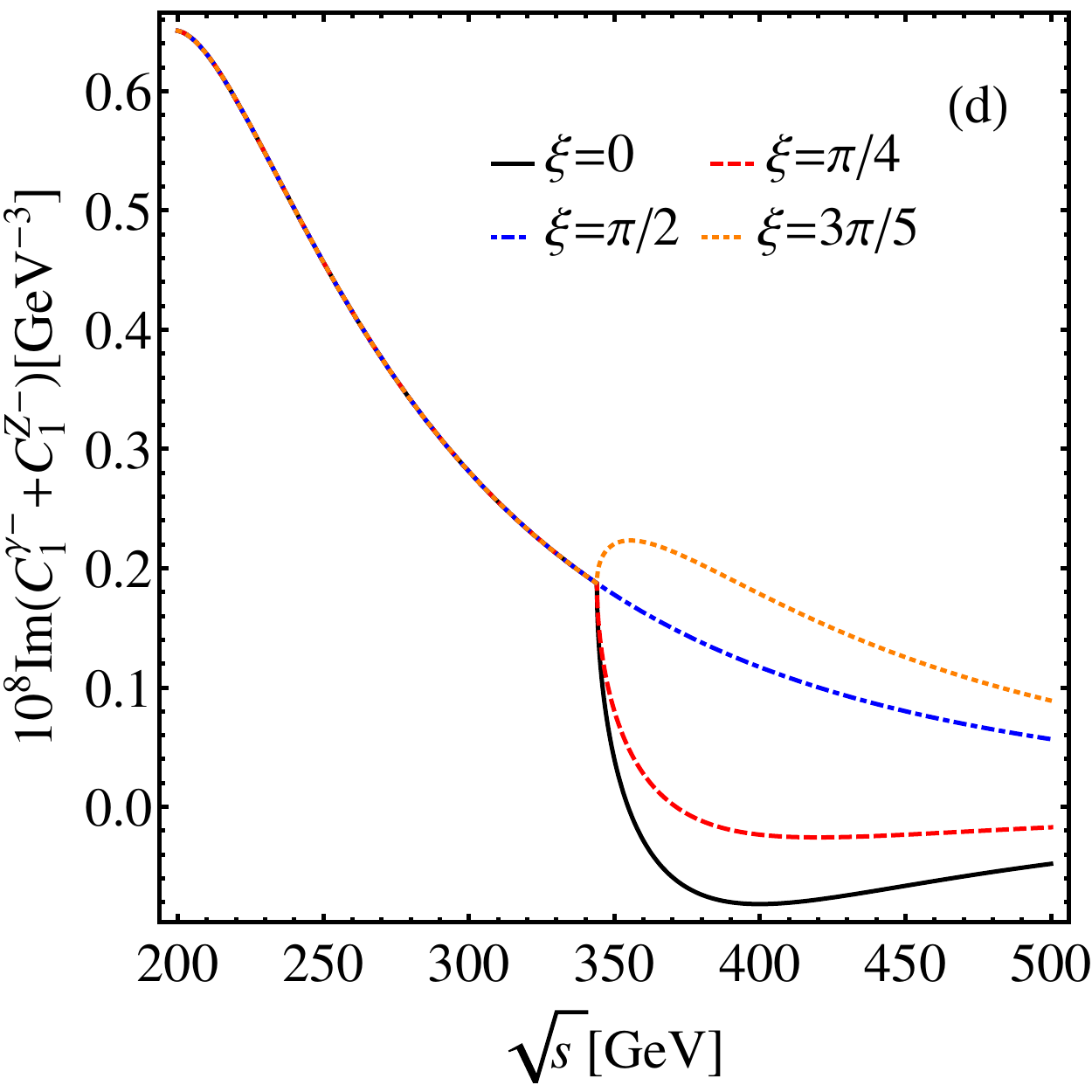}
   \caption{The real and imaginary parts of $C_1^{\gamma\pm}+C_1^{Z\pm}$ for different $\xi$.}.
     \label{fig:c1vertex}
   \end{figure}
 It is apparent that for $2m_W<\sqrt{s}<2m_t$, the imaginary part may arise from $\text{Im}(F_3^{\gamma,W})$ and $\text{Im}(F_3^{Z,W})$, while for $\sqrt{s}>2m_t$, $\text{Im}{F_3^t}$ and $\text{Im}{F_6^t}$ contribute as well. From \EQS{eq:G-F}~\eqref{eq:C-G}, the form factors of s-channel vertex diagrams
 \begin{align}
 C_1^{\gamma\pm}+C_1^{Z\pm}&=-\frac{e}{2s}\frac{e^3m_W}{s_W}[(F_3^{\gamma,W}+\frac{z^{\pm}}{2s_W^2c_W^2(1-m_Z^2/s)}F_3^{Z,W})\nn\\
 &+2Q_tN_c\frac{m_t^2}{m_W^2}(-2Q_t+\frac{z^{\pm}}{2s_W^2c_W^2(1-m_Z^2/s)}g_V^t)F_3^t],
 \end{align} 
where the width of $Z$ boson and light fermion contributions are not included here. In \FIG{fig:c1vertex}, we show the real and imaginary parts of $C_1^{\gamma\pm}+C_1^{Z\pm}$ for different $\xi$. There is a dip at $\sqrt{s}=350\GeV$ for the real part distributions (a)(c) in the SM due to the destructive interference between the $W$ boson and $t$ quark contributions. As $\xi$ increases, the values of real parts become larger since the cancellation is spoiled. For $\xi=\pi/2$, only $W$ boson loops contribute. From (b)(d), the imaginary parts are identical for different $\xi$ as $\sqrt{s}<2m_t$ since only $W$ boson in the loop is on-shell, and differ as $\sqrt{s}$ exceeds the $t\bar{t}$ threshold: for $\xi$ varying from 0 to $3\pi/5$, the imaginary parts increase.

The box diagrams have all partial waves, i.e. $\cos^{n}\theta,n\geq 0$, in general. The form factors $C_1^{\rm{box}\pm}$ and $C_2^{\rm{box}\pm}$ of box diagrams depend on the scattering angle $\theta$ and can provide an imaginary part if the $W,Z$ bosons are on-shell. From \EQ{eq:formfactors} and \FIG{fig:plotbox}, the s-channel vertex diagram contribution interferes with the box diagram contribution destructively in the SM. In the presence of CP-violating $h\bar{t}t$ coupling, the total cross section and differential cross sections for various $\xi$ are discussed in section~\ref{sec:numerical}.

The relation between $\AFB$ and CP violation in this process can be shown via the symmetry of the helicity amplitude $\mathcal{M}_{\lambda,\tau}$, where $\lambda/2\equiv\lambda_{-}=-\lambda_{+}$ with $\lambda_-$ and $\lambda_+$ being the helicities of initial electron and positron respectively, and $\tau=\pm 1$ are the helicities of the final photon. Since we have neglected the electron mass, conservation of the the electron chirality leads that the helicities of initial electron and positron are opposite. Under parity (P) transformation all helicities change signs, while charge conjugation (C) switches $\lambda_{+}$ and $\lambda_{-}$ (exchanges a particle with its antiparticle), that is $\lambda\rightarrow -\lambda$. Thus P, C and CP invariances give rise to~\cite{Jacob:1959at,Gounaris:2015tna}~\footnote{We would like to thank the authors of Ref.~\cite{Gounaris:2015tna} for conversation.}
 \begin{equation}
 \begin{split}
  \mathcal{M}_{\lambda,\tau}(\theta)&=\mathcal{M}_{-\lambda,-\tau}(\theta)\\
 \mathcal{M}_{\lambda,\tau}(\theta)&=\mathcal{M}_{-\lambda,\tau}(\pi-\theta)\\
 \mathcal{M}_{\lambda,\tau}(\theta)&=\mathcal{M}_{\lambda,-\tau}(\pi-\theta)
 \end{split}
 \end{equation}
respectively, up to phases which are however unimportant here~\cite{Farina:2015dua}. For unpolarized beams and the final photon polarizations being summed,
\beq
\sum_{\lambda,\tau=\pm 1}|\mathcal{M}_{\lambda,\tau}|^{2}=|\mathcal{M}_{++}(\cos\theta)|^{2}+|\mathcal{M}_{+-}(\cos\theta)|^{2}+|\mathcal{M}_{-+}(\cos\theta)|^{2}+|\mathcal{M}_{--}(\cos\theta)|^{2}.
\eeq
 P, C and CP invariances imply the relations
 \bea
\sum_{\lambda,\tau=\pm 1}|\mathcal{M}_{\lambda,\tau}|^{2}&=&2(|\mathcal{M}_{++}(\cos\theta)|^{2}+|\mathcal{M}_{+-}(\cos\theta)|^{2}),\\
\label{C}
\sum_{\lambda,\tau=\pm 1}|\mathcal{M}_{\lambda,\tau}|^{2}&=&|\mathcal{M}_{++}(\cos\theta)|^{2}+|\mathcal{M}_{+-}(\cos\theta)|^{2}+|\mathcal{M}_{++}(-\cos\theta)|^{2}+|\mathcal{M}_{+-}(-\cos\theta)|^{2},\\
\label{CP}
\sum_{\lambda,\tau=\pm 1}|\mathcal{M}_{\lambda,\tau}|^{2}&=&|\mathcal{M}_{++}(\cos\theta)|^{2}+|\mathcal{M}_{-+}(-\cos\theta)|^{2}+|\mathcal{M}_{++}(-\cos\theta)|^{2}+|\mathcal{M}_{-+}(\cos\theta)|^{2},
\eea 
respectively. From Eqs.~\eqref{C}~\eqref{CP}, if C and CP are conserved the matrix elements squared are symmetric in $\cos\theta$. Therefore, a nonzero forward-backward asymmetry indicates both C and CP violation. 
 \section{Numerical results}
 \label{sec:numerical}
 In this section, we will give the numerical results of total cross section and differential cross sections at typical c.m. energies of future $e^+e^-$ colliders~\cite{Baer:2013cma}. In practice we insert the Feynman rules of~\EQ{eq:htt} into  FeynArts-3.9~\cite{Hahn:2000kx} model file, and calculate the amplitude automatically using FormCalc-8.4~\cite{Hahn:1998yk} and LoopTools-2.12~\cite{Hahn:1998yk}. We have checked both analytically and numerically the ultraviolet (UV) finiteness of the amplitudes.

\begin{figure}
  \centering
 \includegraphics[width=0.4\textwidth]{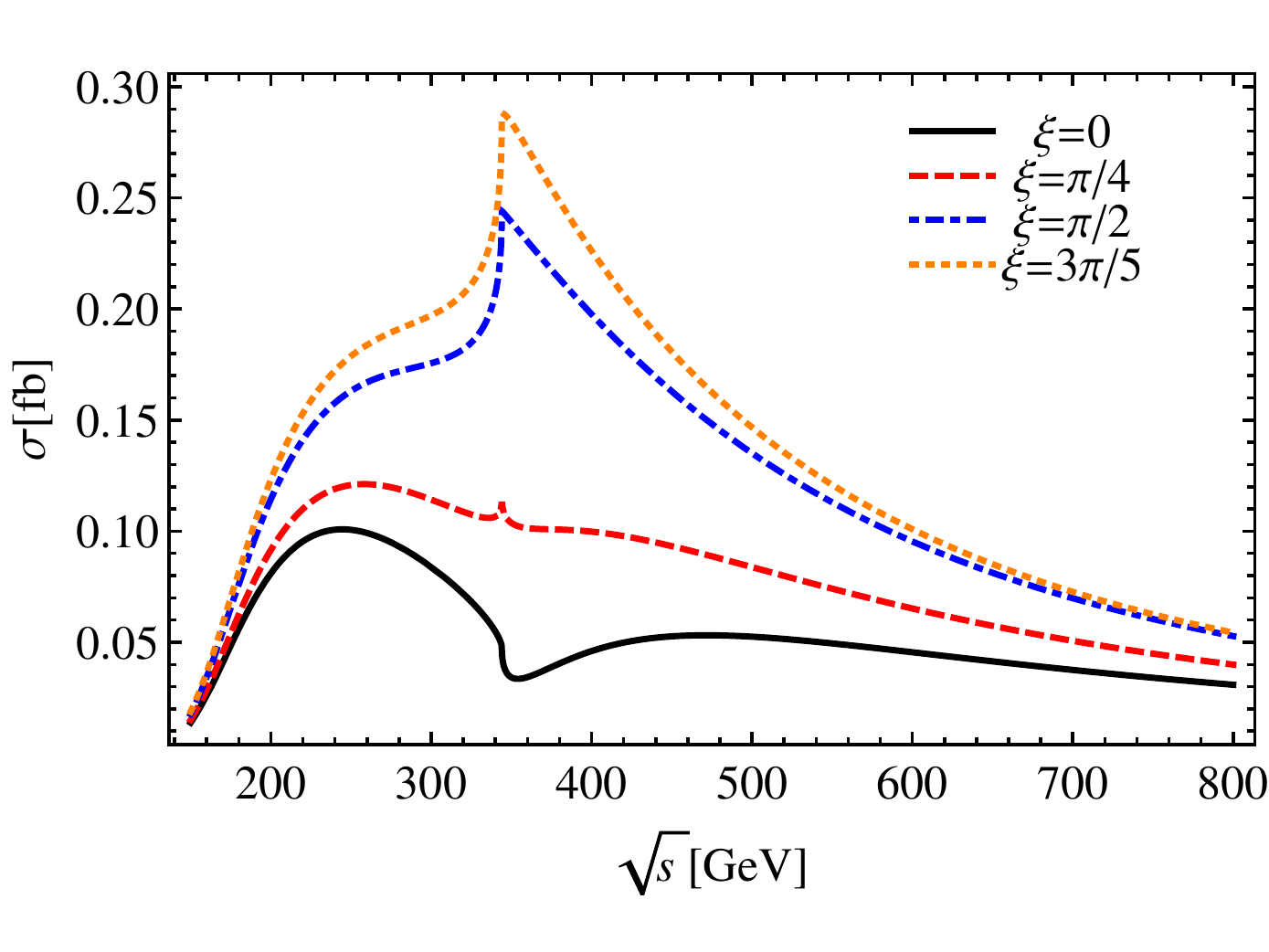}
 \includegraphics[width=0.4\textwidth]{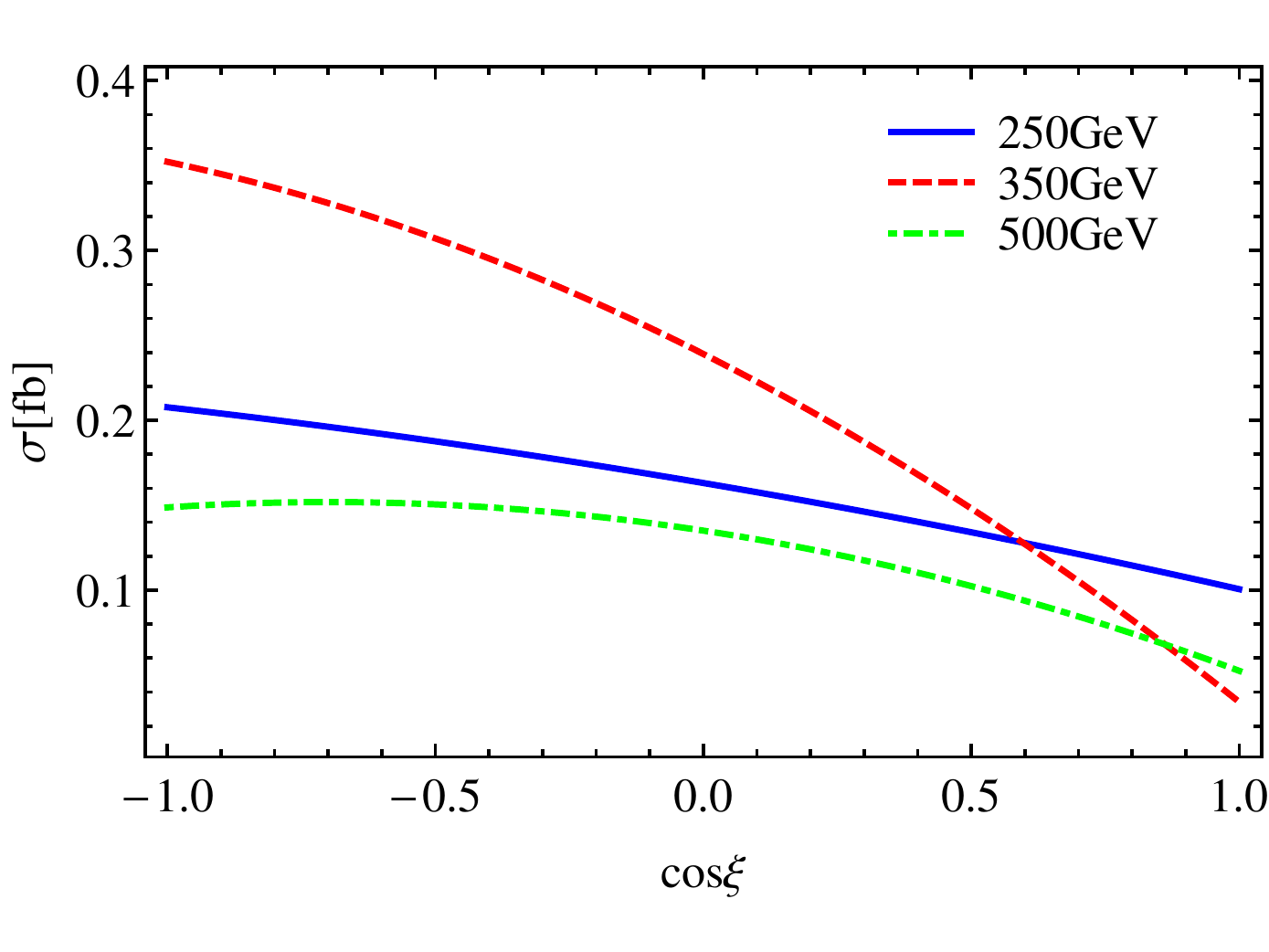}
   \caption{Left: cross sections as a function of c.m. energy $\sqrt{s}$  for the CP phases $\xi=0$ (black, solid), $\pi/4$ (red, dashed), $\pi/2$ (blue, dotdashed), $3\pi/5$ (orange, dotted). Right: cross sections as a function of $\cos{\xi}$ for the c.m. energies $\sqrt{s}=250\GeV$ (blue, solid), $350\GeV$ (red, dashed), $500\GeV$ (green, dotdashed).}
 \label{fig:xsex}
   \end{figure}

\FIG{fig:xsex} shows the cross sections of $e^+e^-\to h\gamma$ for different c.m. energies $\sqrt{s}$ and CP phases $\xi$. From the left plot, we can see that the cross section grows as $\xi$ increases from 0 to $3\pi/5$. The shift of c.m. energy corresponding to the maximal value of the cross section from $250\GeV$ to $350\GeV$ is due to the competition of the $W$-loop functions with the $t$-loop function of the s-channel diagrams and their interference with the box  diagram contribution. From the right plot, we can see that the cross section at $\sqrt{s}=350~\rm{GeV}$ drops rapidly ($\sim 10$ times) as $|\xi|$ decreases. Thus measuring the cross section of this process at $350~\rm{GeV}$ can be helpful to search for CP violation in the $ht\bar{t}$ interaction.
 \begin{figure}
  \centering
 \includegraphics[width=0.4\textwidth]{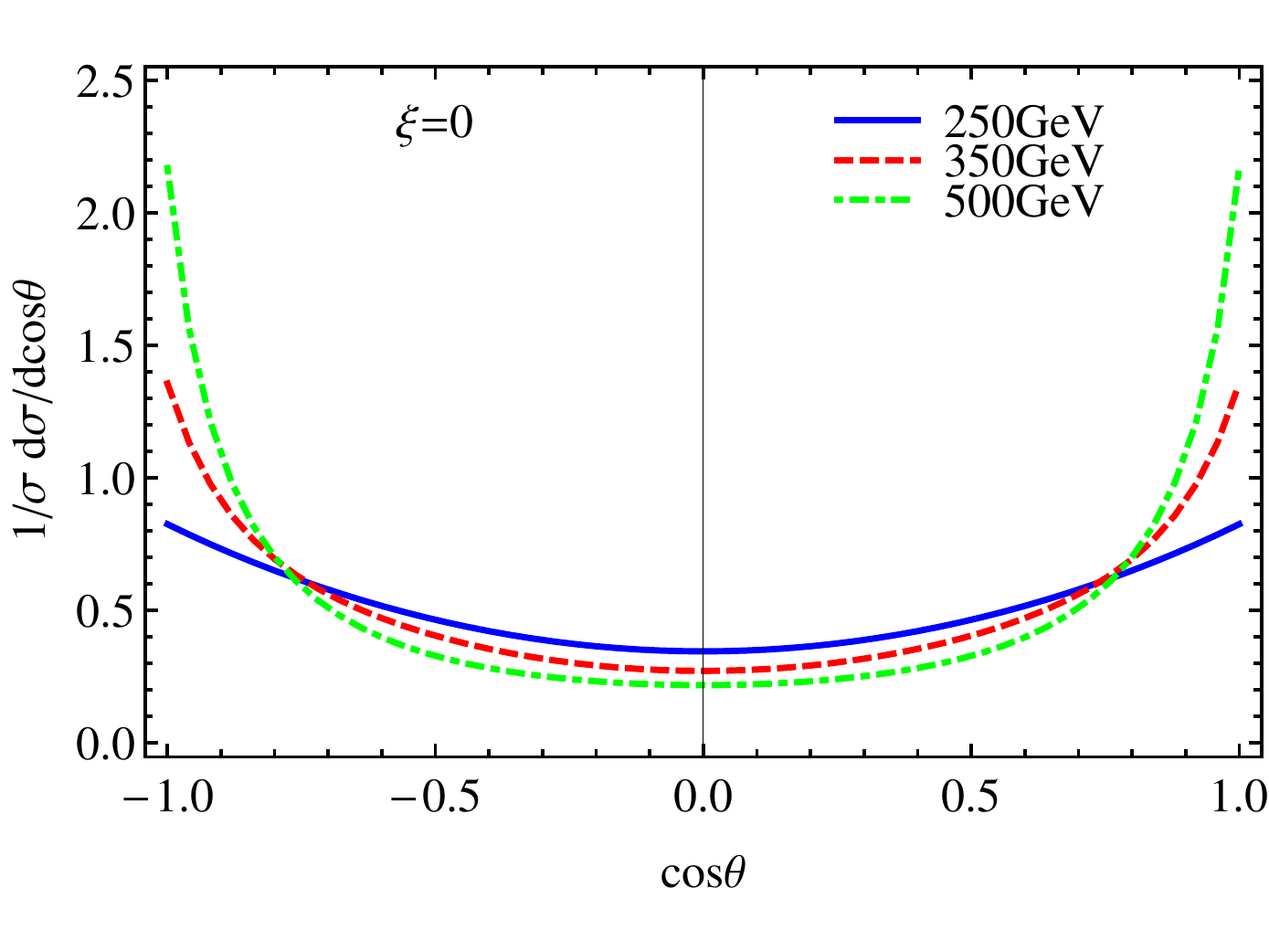}
 \includegraphics[width=0.4\textwidth]{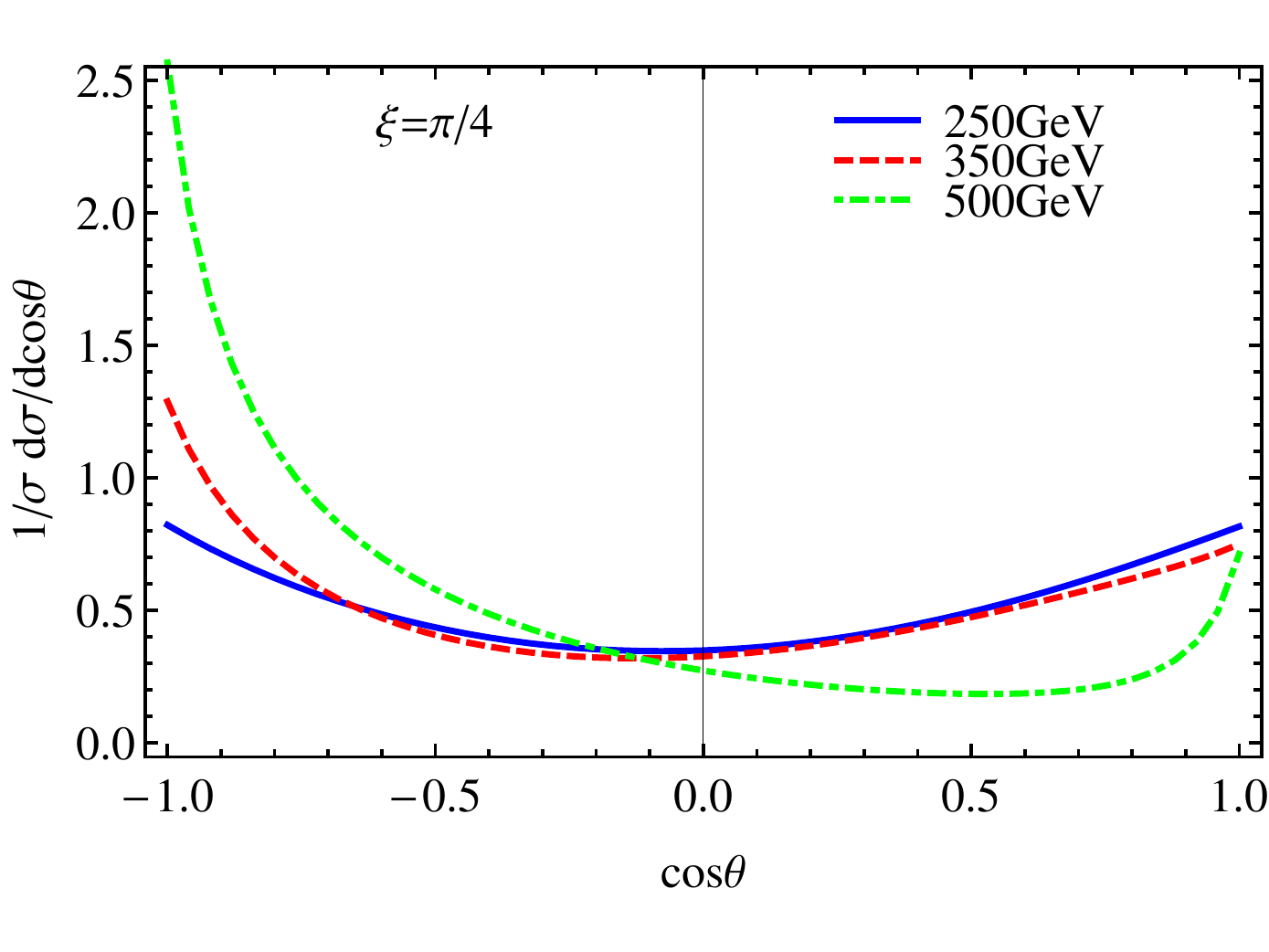}\\
 \includegraphics[width=0.4\textwidth]{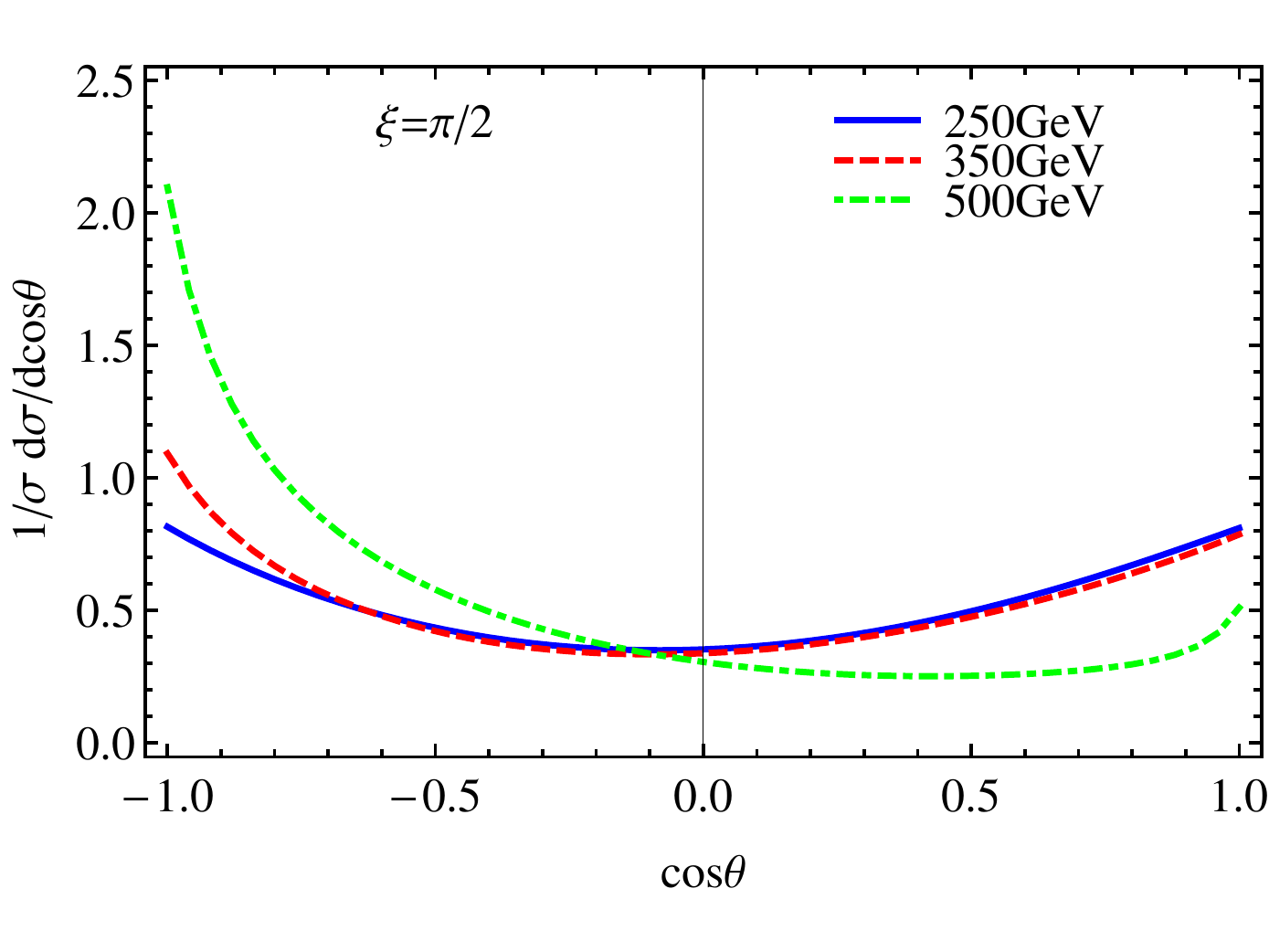}
 \includegraphics[width=0.4\textwidth]{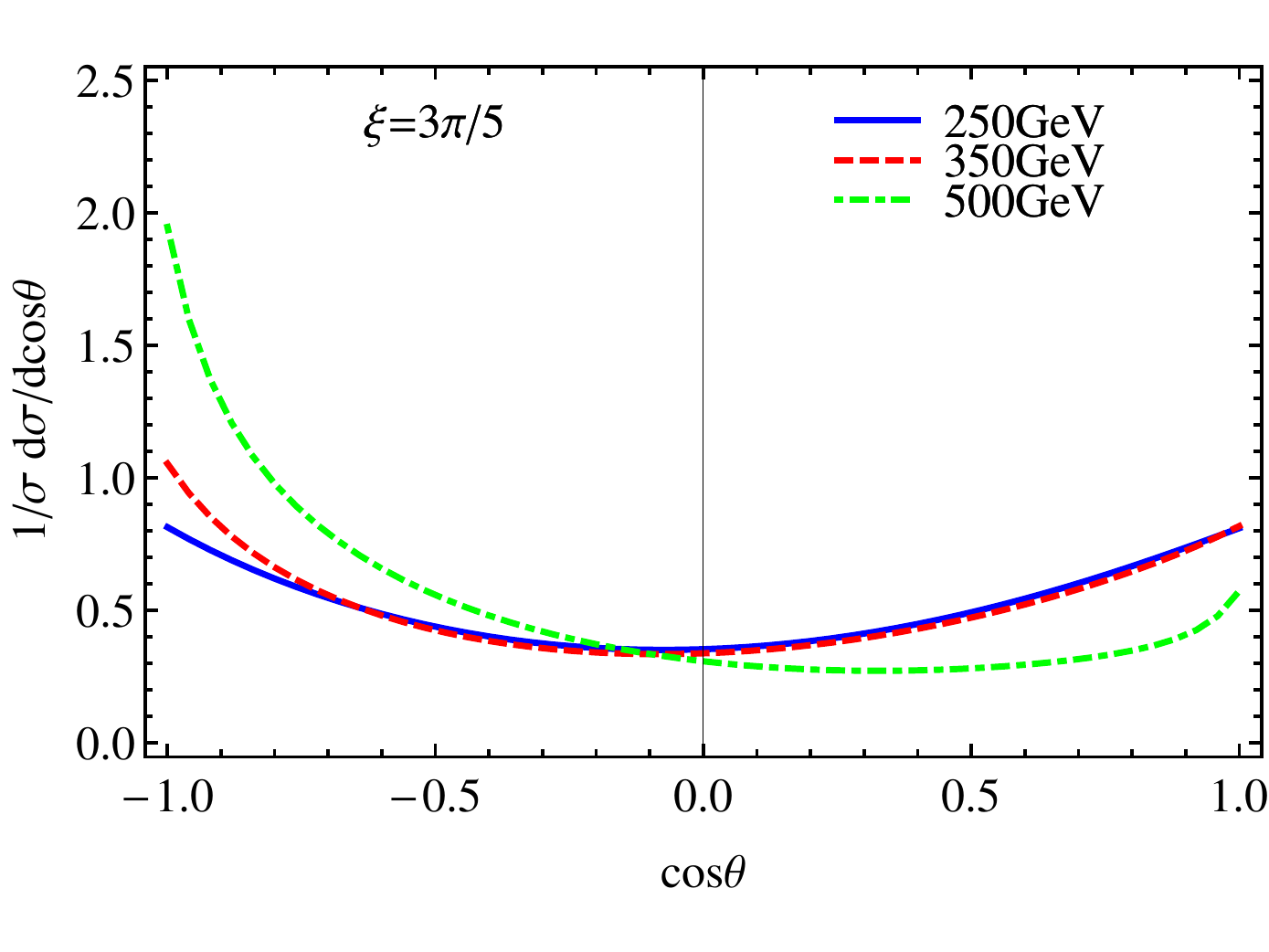}
   \caption{Normalized distributions of the scattering angle $\theta$ for different $\xi$ and $\sqrt{s}$. $\theta$ is defined as the polar angle between the momenta of final photon and positron. For $\xi=0$, the distribution is symmetric. For $\xi=\pi/2,\ \xi=\pi/4,\ 3\pi/5$, the distributions are asymmetric.}
 \label{fig:dxsex}
   \end{figure}  
 \begin{figure}
  \centering
     \includegraphics[width=0.3\textwidth]{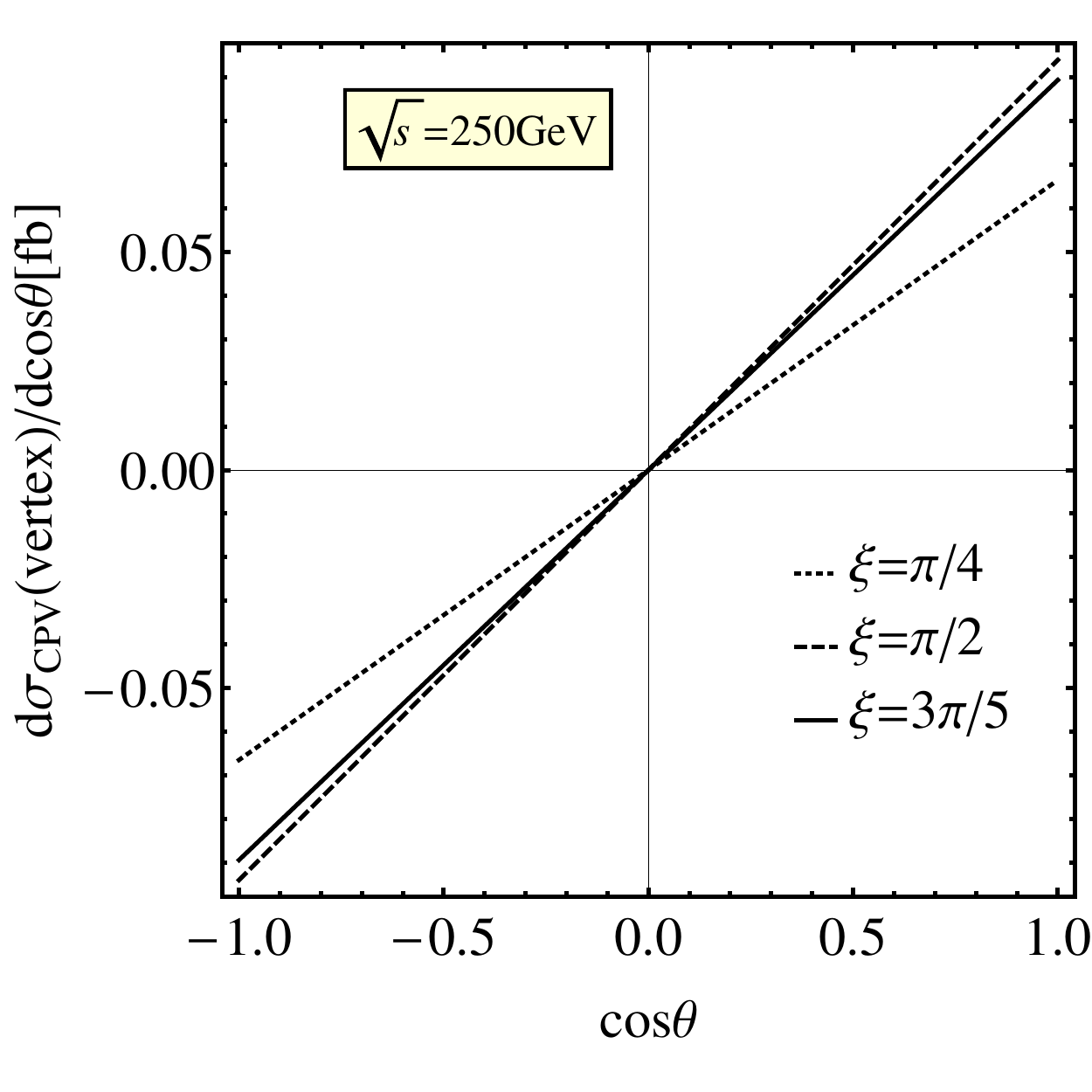}
  \includegraphics[width=0.3\textwidth]{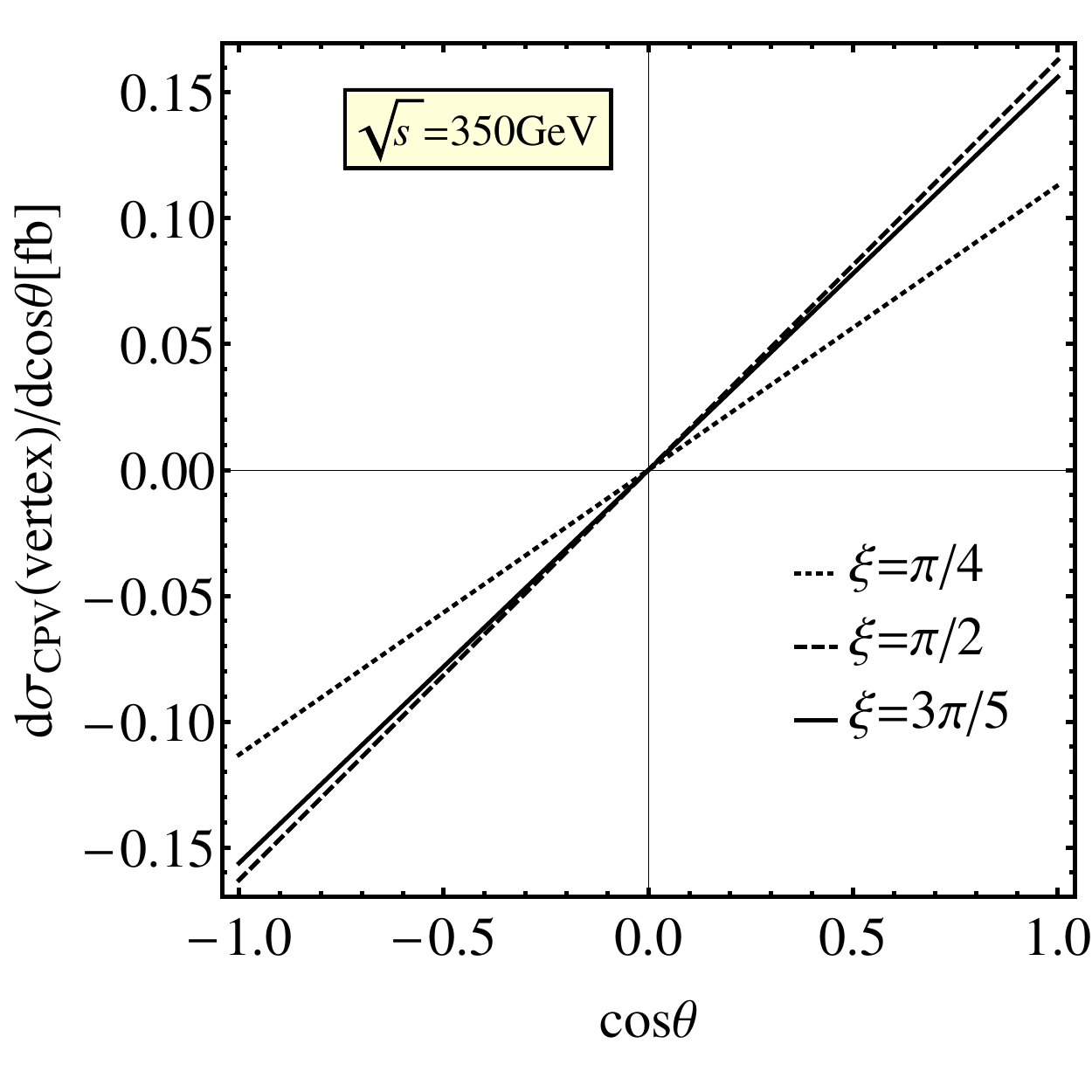}
  \includegraphics[width=0.3\textwidth]{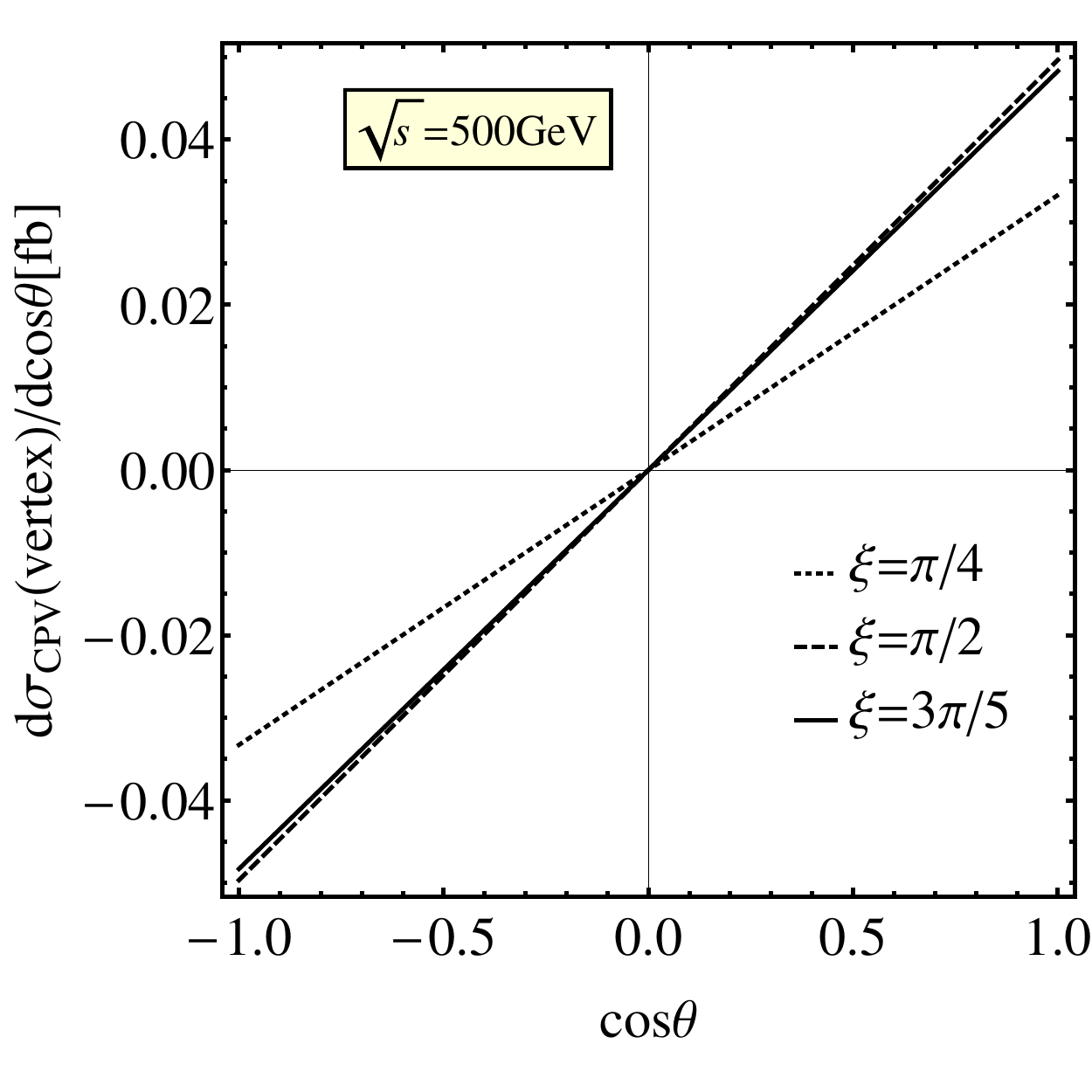}
   \includegraphics[width=0.3\textwidth]{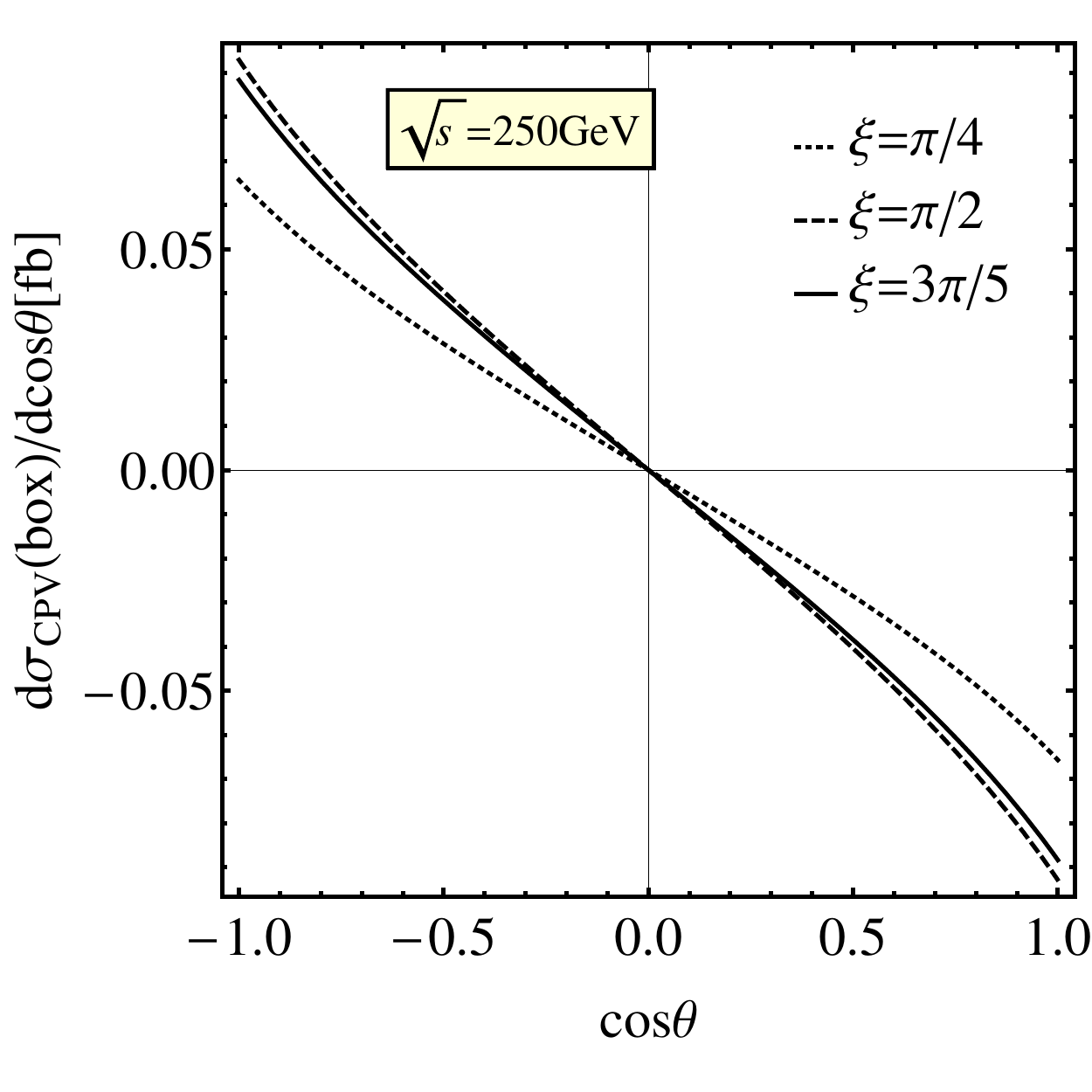}
  \includegraphics[width=0.3\textwidth]{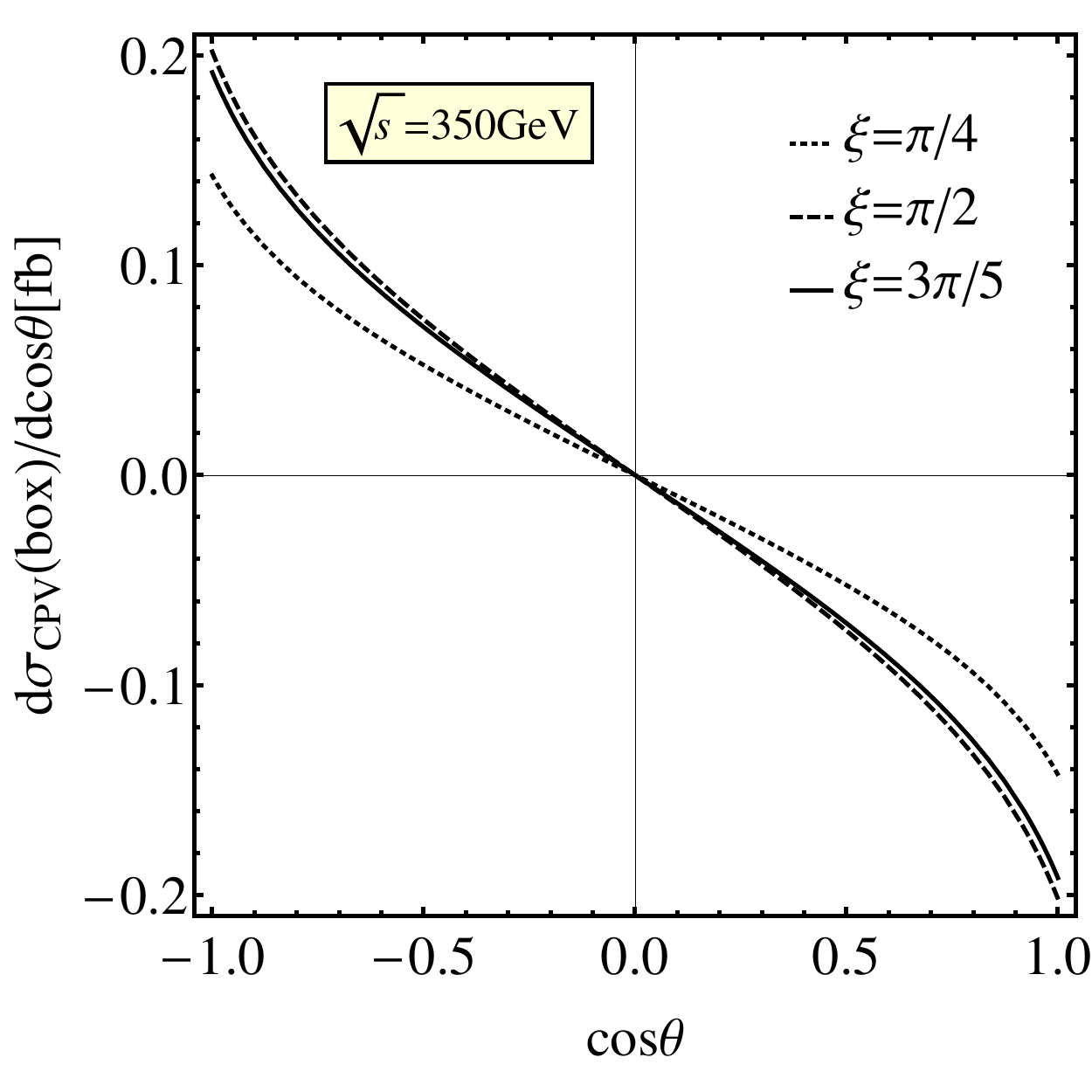}
  \includegraphics[width=0.3\textwidth]{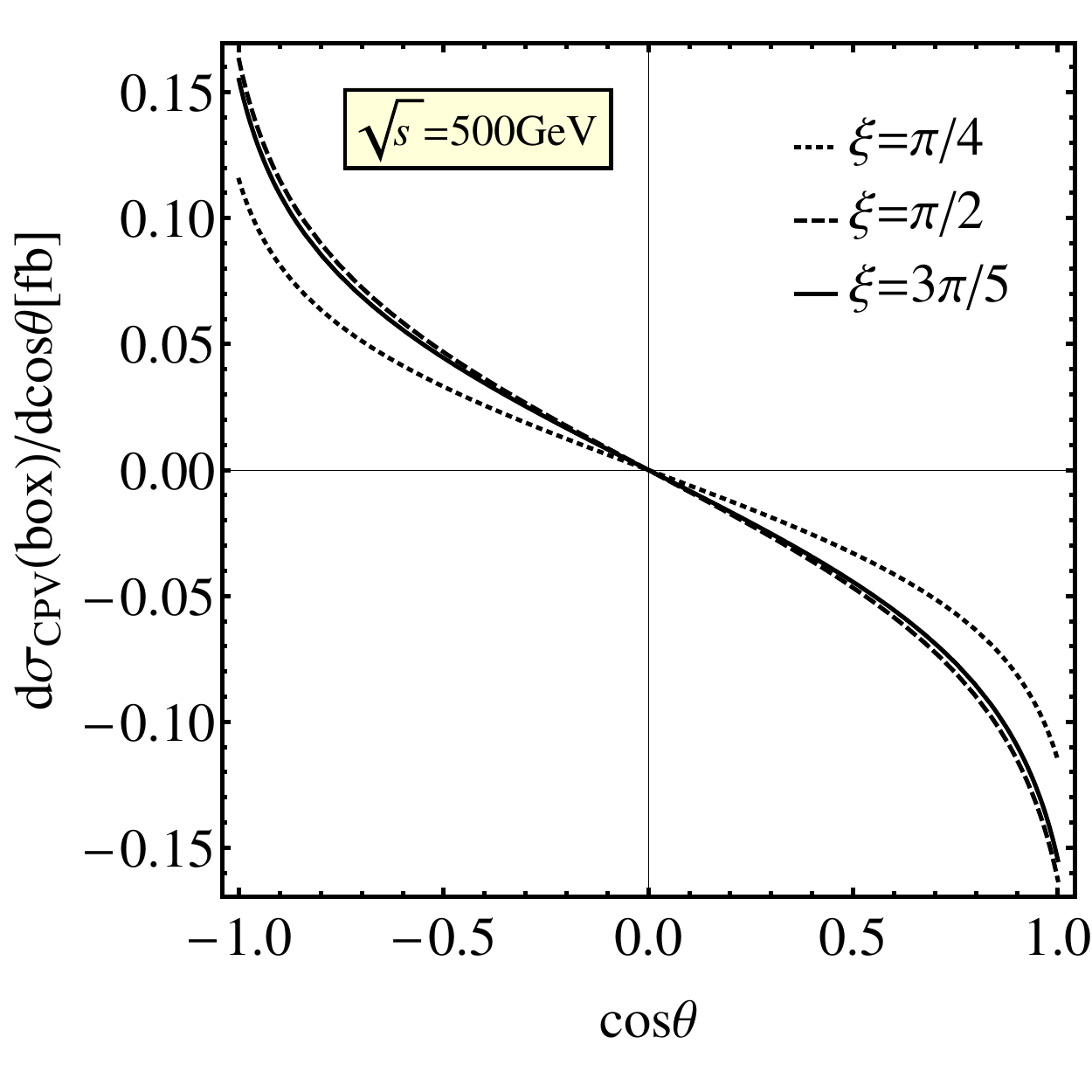} 
 \includegraphics[width=0.3\textwidth]{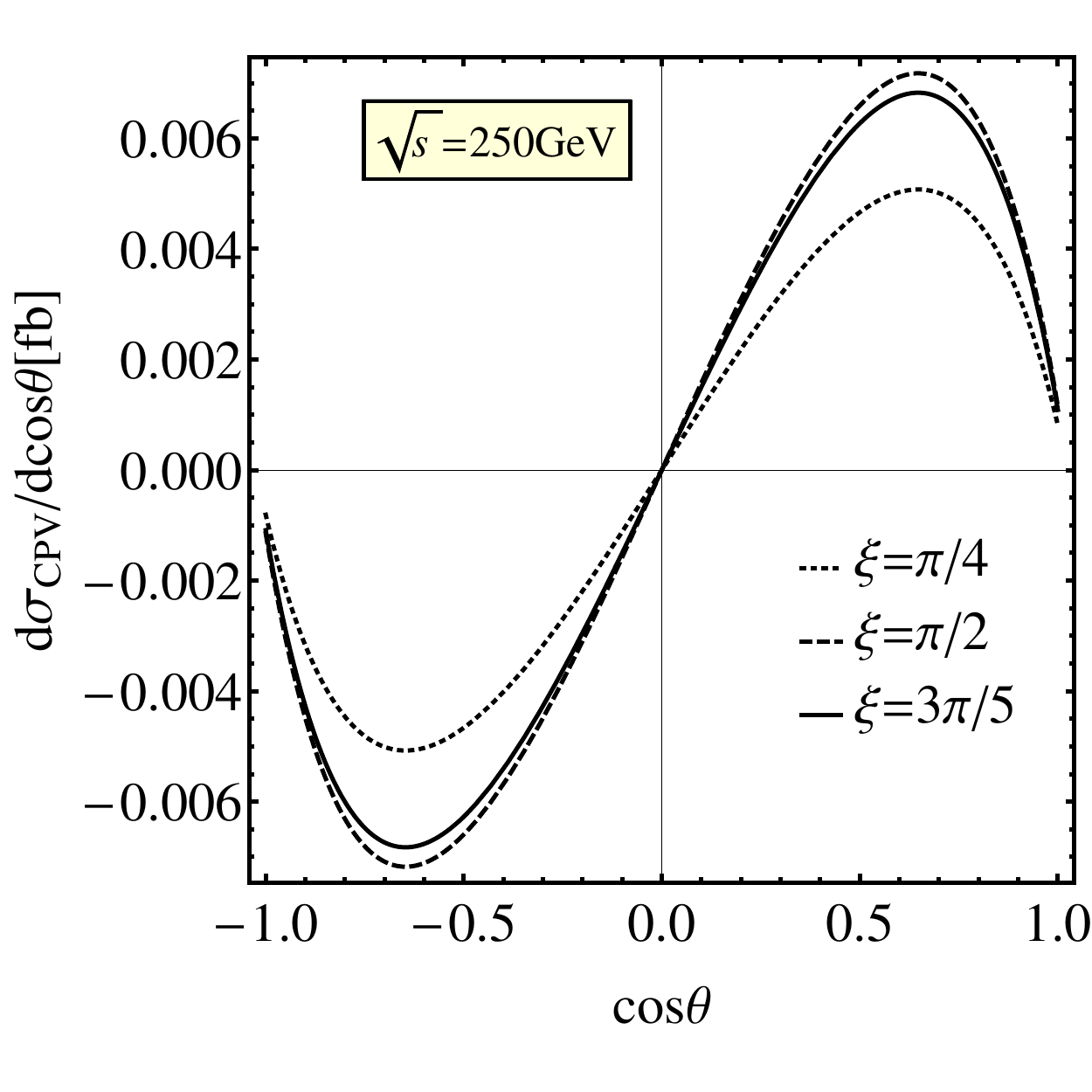}
  \includegraphics[width=0.3\textwidth]{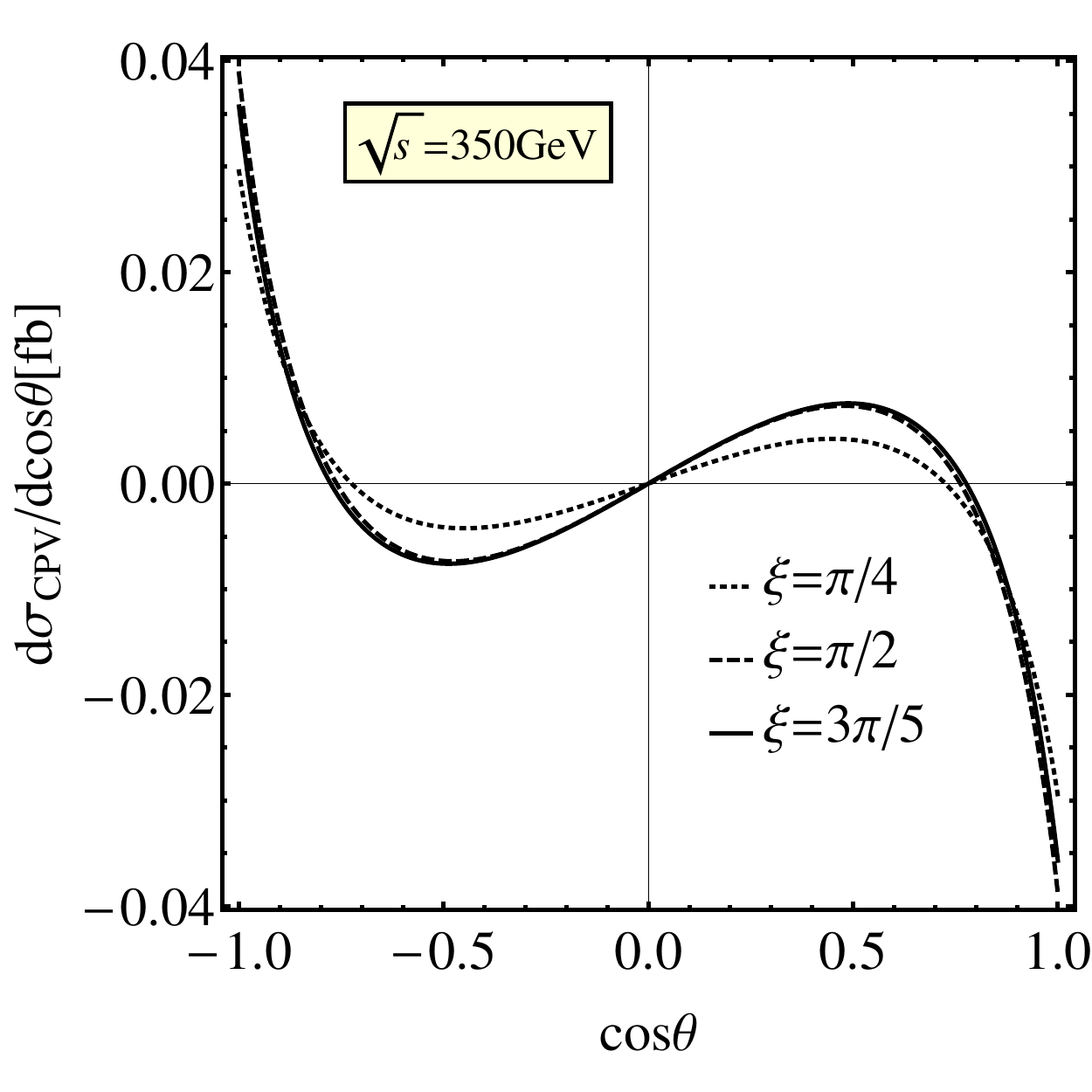}
  \includegraphics[width=0.3\textwidth]{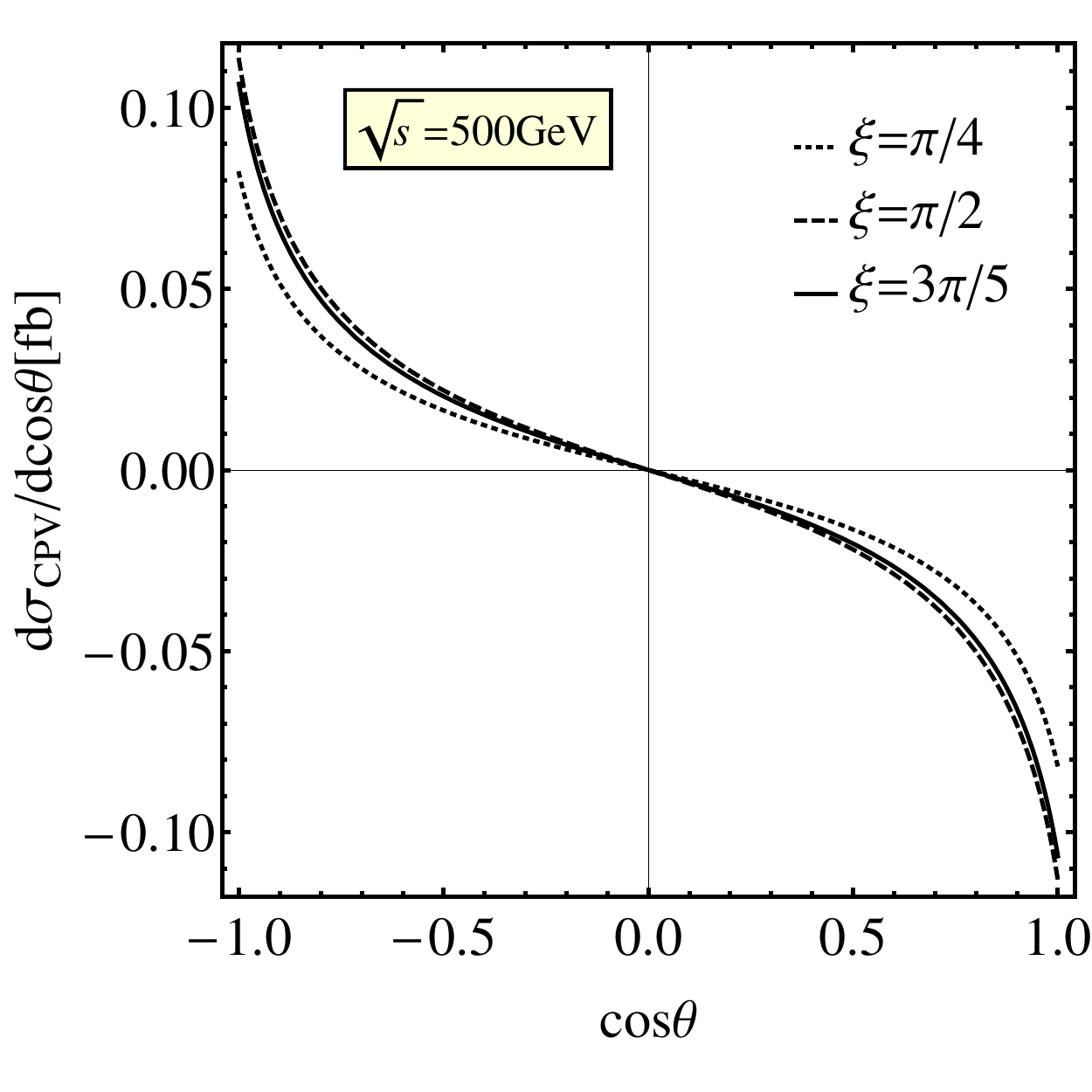}
   \caption{The distributions of $d\sigma_{CPV}(\rm{vertex})/d\cos\theta$, $d\sigma_{CPV}(\rm{box})/d\cos\theta$ and $d\sigma_{CPV}/d\cos\theta$ at c.m. energies $\sqrt{s}=250\GeV$, $250\GeV$ and $500\GeV$, where we have not included the light fermion contributions. The dotted, dashed and solid curves correspond to $\xi=\pi/4,\pi/2,3\pi/5$, respectively.}
 \label{fig:interference}
   \end{figure}   
Besides the total cross section, the differential cross section is also important. \FIG{fig:dxsex} shows the normalized angular distributions for different $\sqrt{s}$ and $\xi$. The distribution is symmetric in the SM ($\xi=0$) and becomes asymmetric in the forward ($\cos\theta>0$) and backward ($\cos\theta<0$) regions as $\xi$ deviates from zero, even in the case of $\xi=\pi/2$~\footnote{This depends on our assumption of the couplings $hWW$ and $hZZ$. In the 2HDMs~\cite{Branco:2011iw}, however, the CP-odd Higgs does not couple to the gauge bosons $W,Z$ at tree-level.}. Since the coefficients $\mathcal{F}_{1\pm}$ and $\mathcal{F}_{2\pm}$ in \EQ{cpv2} depend on the scattering angle $\theta$, the distributions for $\xi\neq 0$ are not parabolic in shape. The asymmetry is most apparent at $\sqrt{s}=500\GeV$.
   
Concentrate on the fist two terms in \EQ{cpv1} and 
define the "differential cross sections":
\bea
\frac{d\sigma_{CPV}(\rm{ver})}{d\cos\theta}&=&K\{[(C_{6}^{-*}C_{1}^{\rm{ver}-}-C_{6}^{-}C_{1}^{\rm{ver}-*})-(C_{6}^{+*}C_{1}^{\rm{ver}+}-C_{6}^{+}C_{1}^{\rm{ver}+*})](1+\cos\theta)^2\nn\\
&-&[(C_{6}^{-*}C_{2}^{\rm{ver}-}-C_{6}^{-}C_{2}^{\rm{ver}-*})-(C_{6}^{+*}C_{2}^{\rm{ver}+}-C_{6}^{+}C_{2}^{\rm{ver}+*})](1-\cos\theta)^2\},\\
\frac{d\sigma_{CPV}(\rm{box})}{d\cos\theta}&=&K\{[(C_{6}^{-*}C_{1}^{\rm{box}-}-C_{6}^{-}C_{1}^{\rm{box}-*})-(C_{6}^{+*}C_{1}^{\rm{box}+}-C_{6}^{+}C_{1}^{\rm{box}+*})](1+\cos\theta)^2\nn\\
&-&[(C_{6}^{-*}C_{2}^{\rm{box}-}-C_{6}^{-}C_{2}^{-*})-(C_{6}^{+*}C_{2}^{\rm{box}+}-C_{6}^{+}C_{2}^{\rm{box}+*})](1-\cos\theta)^2\},\\
\frac{d\sigma_{CPV}}{d\cos\theta}&=&K\{[(C_{6}^{-*}C_{1}^{-}-C_{6}^{-}C_{1}^{-*})-(C_{6}^{+*}C_{1}^{+}-C_{6}^{+}C_{1}^{+*})](1+\cos\theta)^2\nn\\
&-&[(C_{6}^{-*}C_{2}^{-}-C_{6}^{-}C_{2}^{-*})-(C_{6}^{+*}C_{2}^{+}-C_{6}^{+}C_{2}^{+*})](1-\cos\theta)^2\},
\eea
with $C_{1,2}^{\rm{ver}\pm}\equiv C_{1,2}^{\gamma\pm}+C_{1,2}^{Z\pm}$, $C_{1,2}^{\pm}=C_{1,2}^{\rm{ver}\pm}+C_{1,2}^{\rm{box}\pm}$ and the overall factor $K=i(s-m_h^2)^3/(4^8\pi^5 s)$.
For the s-channel vertex diagrams $C_{1}^{\rm{ver}\pm}=C_{2}^{\rm{ver}\pm}$, thus $d\sigma_{CPV}(\rm{ver})/d\cos\theta$ is proportional to $\cos\theta$. For the box diagrams, the dependence on $\cos\theta$ is not manifest since the expressions of $C_{1}^{\rm{box}\pm}$ and $C_{2}^{\rm{box}\pm}$ are complicated as displayed in Appendix~\ref{loopfunctions}. In \FIG{fig:interference}, we show $d\sigma_{CPV}(\rm{vertex})/d\cos\theta$, $d\sigma_{CPV}(\rm{box})/d\cos\theta$ and $d\sigma_{CPV}/d\cos\theta$ numerically for different CP phases $\xi$ at $\sqrt{s}=250\GeV,350\GeV,500\GeV$. From the plots in the first two rows, the s-channel vertex diagram contribution and box diagram contribution to the difference of cross sections in the forward and backward regions, i.e. $\sigma_F-\sigma_B$ are destructive. Furthermore, the magnitudes of both are maximal for $\xi=\pi/2$, since $C_6^{\pm}$ are proportional to $\sin\xi$.
At $\sqrt{s}=250\GeV$, the magnitudes of $d\sigma_{CPV}(\rm{ver})/d\cos\theta$ and $d\sigma_{CPV}(\rm{box})/d\cos\theta$ are comparable so that there is a large cancellation between them and the resulting differential cross section $d\sigma_{CPV}/d\cos\theta$ is small. Besides, $\sigma_F-\sigma_B$ at $\sqrt{s}=250\GeV$ is always positive. At $\sqrt{s}=350\GeV$, both $d\sigma_{CPV}(\rm{vertex})/d\cos\theta$ and $d\sigma_{CPV}(\rm{box})/d\cos\theta$ are enhanced, since the loop function $C_6^{\pm}$ gets its largest value near the $t\bar{t}$ threshold. The resulting distribution of $d\sigma_{CPV}/d\cos\theta$ is not monotonic and the difference $\sigma_F-\sigma_B$ is positive for $\xi=3\pi/5$ while negative for $\xi=\pi/4,\pi/2$ after integrating $d\sigma_{CPV}/d\cos\theta$ in the forward and backward regions. The magnitude of the box diagram contribution at $\sqrt{s}=500\GeV$ is about 3 times larger than the box diagram contribution at $\sqrt{s}=250\GeV$ while the s-channel vertex diagram contribution at $\sqrt{s}=500\GeV$ is smaller. As a result, $d\sigma_{CPV}/d\cos\theta$ at $\sqrt{s}=500\GeV$ is two orders of magnitude larger and the difference $\sigma_F-\sigma_B$ is negative.

\begin{figure}[!htb]
  \centering
 \includegraphics[width=0.4\textwidth]{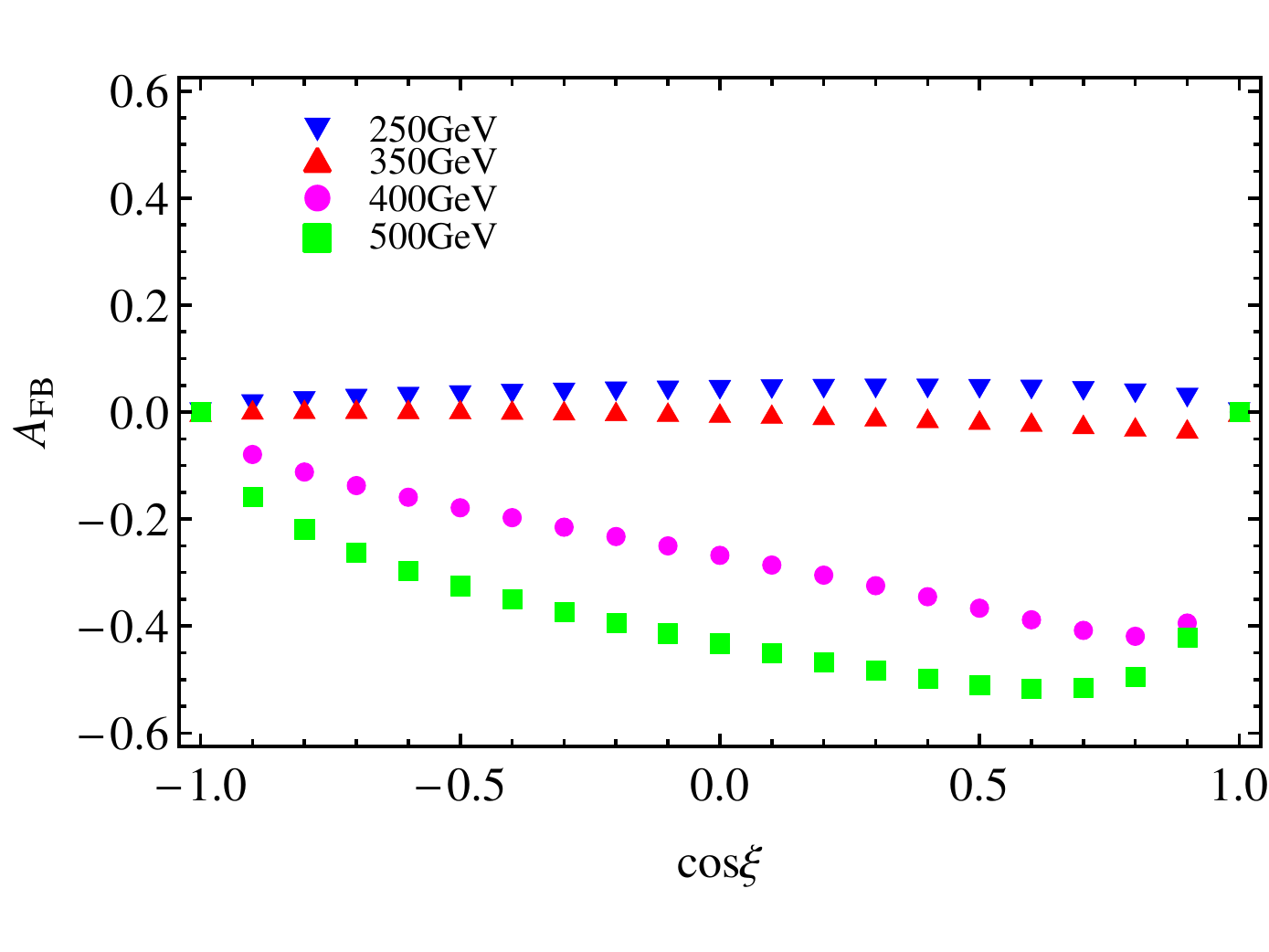}
   \caption{$\AFB$ generated for various CP phases $\xi$ and c.m. energies $\sqrt{s}=250\GeV$ (blue, triangle), $350\GeV$ (red, triangle), $400\GeV$ (purple, disk), $500\GeV$ (green, rectangle).}
     \label{fig:afb}
   \end{figure}

To illustrate the asymmetry quantitatively, we present the values of $A_{FB}$ in \FIG{fig:afb} and Table~\ref{tab:afb} ($A_{FB}$ before cuts are applied) given the c.m. energy $\sqrt{s}$ and CP phase $\xi$. The $A_{FB}$ depends on both the difference $\sigma_F-\sigma_B$ and the total cross section, so $A_{FB}$ and $\sigma_F-\sigma_B$ have the same sign. We find that the magnitude of $A_{FB}$ \textit{tends} to be small at low $\sqrt{s}$ and get larger as $\sqrt{s}$ increases\footnote{For illustration, we also show $A_{FB}$ at $\sqrt{s}=400\GeV$ for varying $\cos\xi$ in \FIG{fig:afb}.}. In Table~\ref{tab:afb} ($A_{FB}$ before cuts are applied), the $A_{FB}$ is positive at $\sqrt{s}=250\GeV$ but negative at $\sqrt{s}=500\GeV$, and  the latter is two orders of magnitude larger since $\sigma_F-\sigma_B$ at $\sqrt{s}=500\GeV$ is two orders of magnitude larger while their total cross sections are comparable, see \FIG{fig:xsex}. Furthermore, the $A_{FB}$ at $\sqrt{s}=350\GeV$ with $\xi=\pi/2,3\pi/5$ are tiny in consideration of the small $\sigma_F-\sigma_B$ and the enhancement of total cross sections.
\section{Collider simulation}
\label{sec:collider}
In this section, we will simulate the signal process $\eebbah$ and its backgrounds at the future high luminosity ($\mathcal{L}=1~\text{ab}^{-1}$, $3~\text{ab}^{-1}$, $10~\text{ab}^{-1}$)  $e^{-}e^{+}$ colliders at the typical c.m. energy $\sqrt{s}=250\GeV$, $350\GeV$, $500\GeV$. To generate signal events, we obtain the amplitude squared for $\eeha$ with the help of
FeynArts~\cite{Hahn:2000kx} and FormCalc/LoopTools packages~\cite{Hahn:1998yk} and then pass them to MadGraph~\cite{Alwall:2011uj, Alwall:2014hca}. 
The background matrix element for $e^{+}e^{-}\rightarrow b\bar{b}\gamma$ is generated directly using MadGraph.  The entire analysis is done at the parton level, with the following event selection cuts being applied:
\bea
\label{selectcut1}
 (a)&~p^{b,\bar{b}}_{T}> 20\GeV,\ p^{\gamma}_{T}>25\GeV,\ |\eta^{b,\bar{b},\gamma}|\leq 2.5,\ \Delta R_{b\bar{b},b\gamma,\bar{b}\gamma}\geq 0.4,\\
 \label{selectcut2}
(b)&~|m_{b\bar{b}}-m_{h}|\leq 15\GeV,\ \Delta E_{\gamma}< E_{\gamma}\times 0.5\%
\eea
where $p_{T}^{i}$ and $\eta^{i}$ denote the transverse momentum and pseudo-rapidity of the particle $i$, respectively. The spatial separation  between the objects $k$ and $l$ is denoted by $\Delta R_{kl}$. The photon in the signal event exhibits a harder transverse momentum to balance the momentum of Higgs boson than the photon in the background which is mainly radiated from the initial electron and positron and peaks in the small $p_{T}$ owing to the collinear enhancement. The $b$-tagging efficiency $\epsilon_{b}$ and the mis-tag probabilities $\epsilon_{j\rightarrow b}$ for light jets in our analysis are~\cite{Behnke:2013lya}
\beq
\label{efficiencies}
\epsilon_{b}=0.9,\ \epsilon_{c\rightarrow b}=0.1,\ \epsilon_{j\rightarrow b}\simeq 0,\ \text{for}\ j=u,d,s,g,
\eeq
and at least 1 $b$-jet is tagged. 
 \begin{figure}
  \centering
\includegraphics[width=0.4\textwidth]{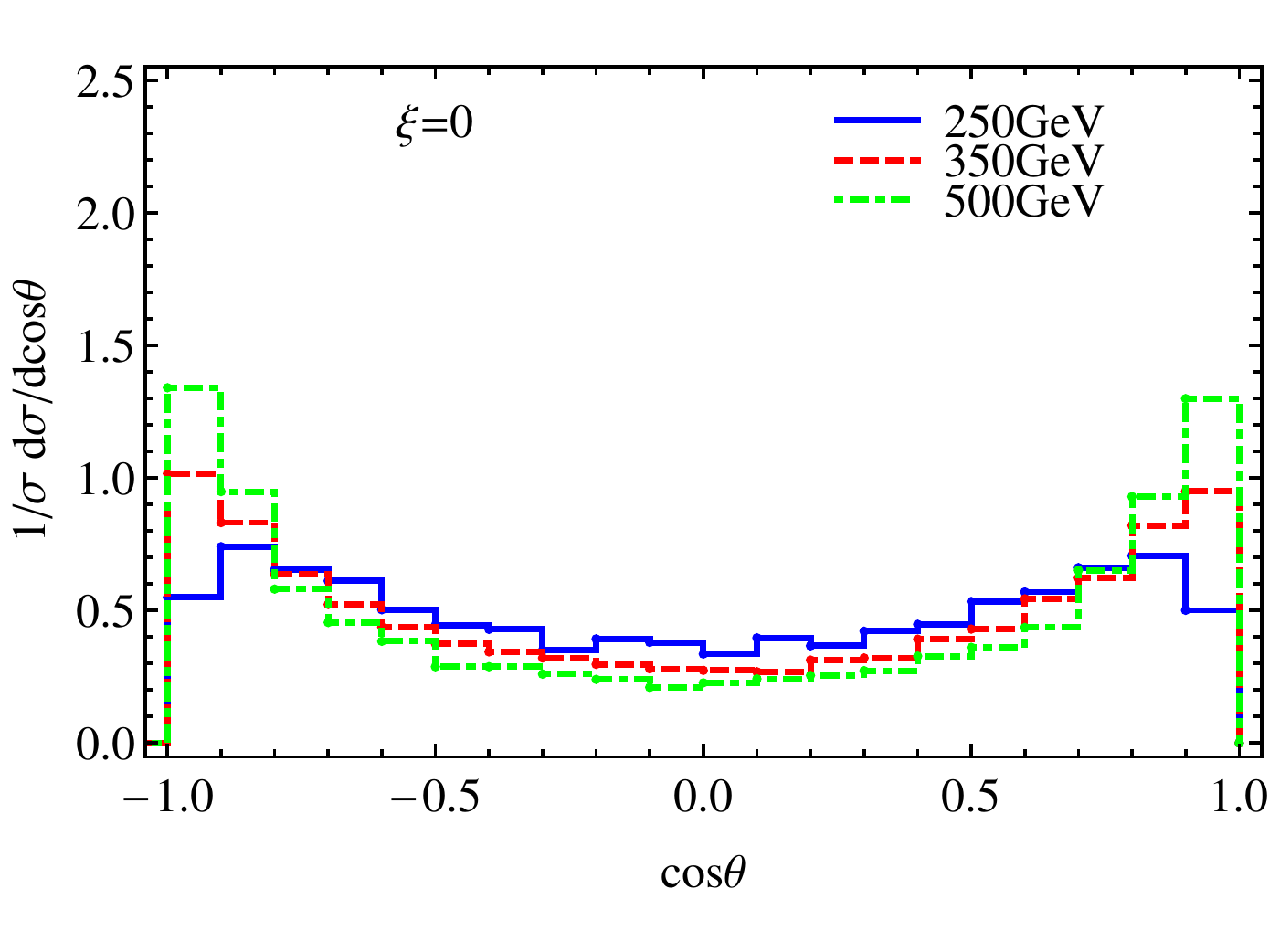}
\includegraphics[width=0.4\textwidth]{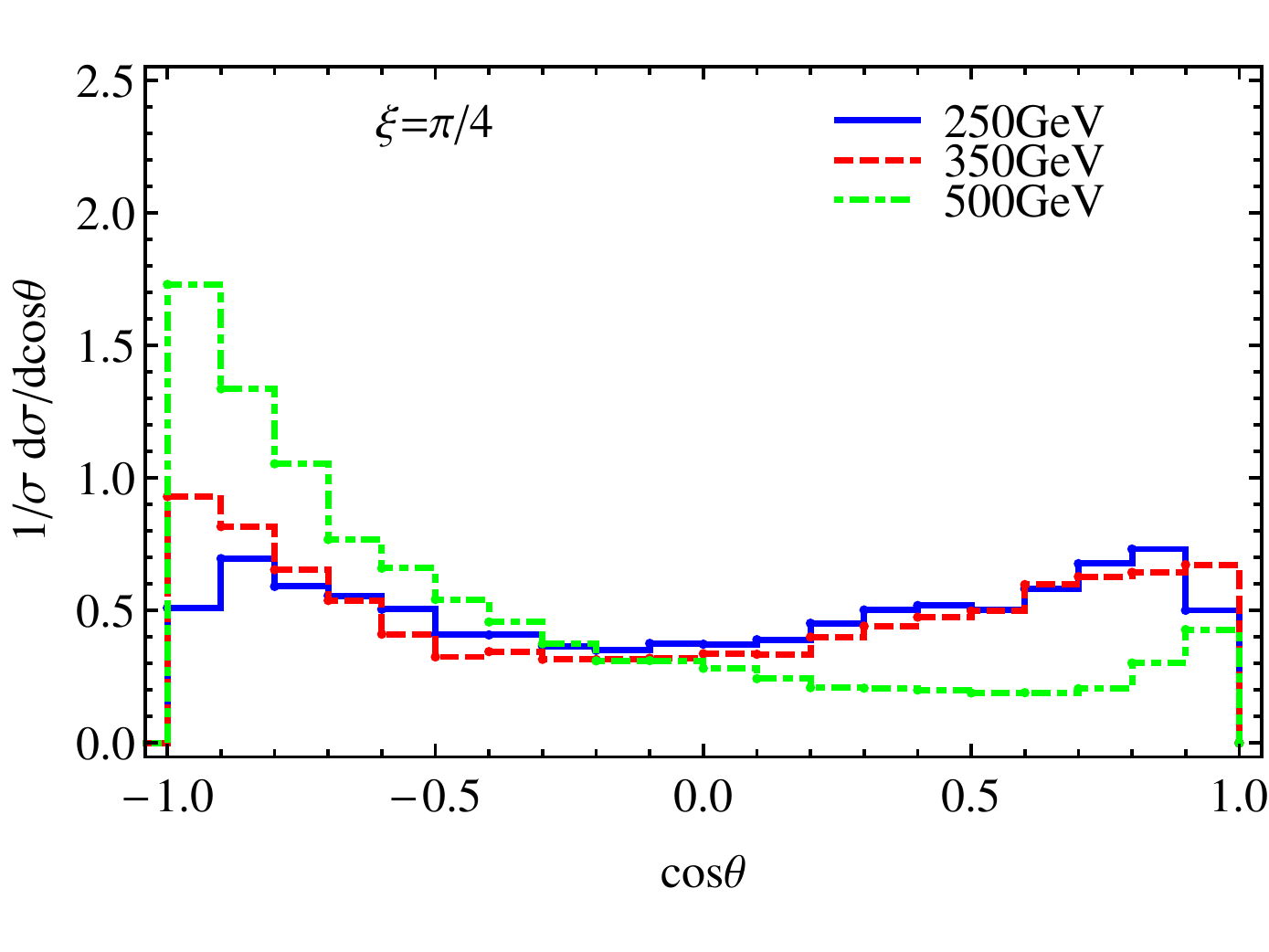}\\
\includegraphics[width=0.4\textwidth]{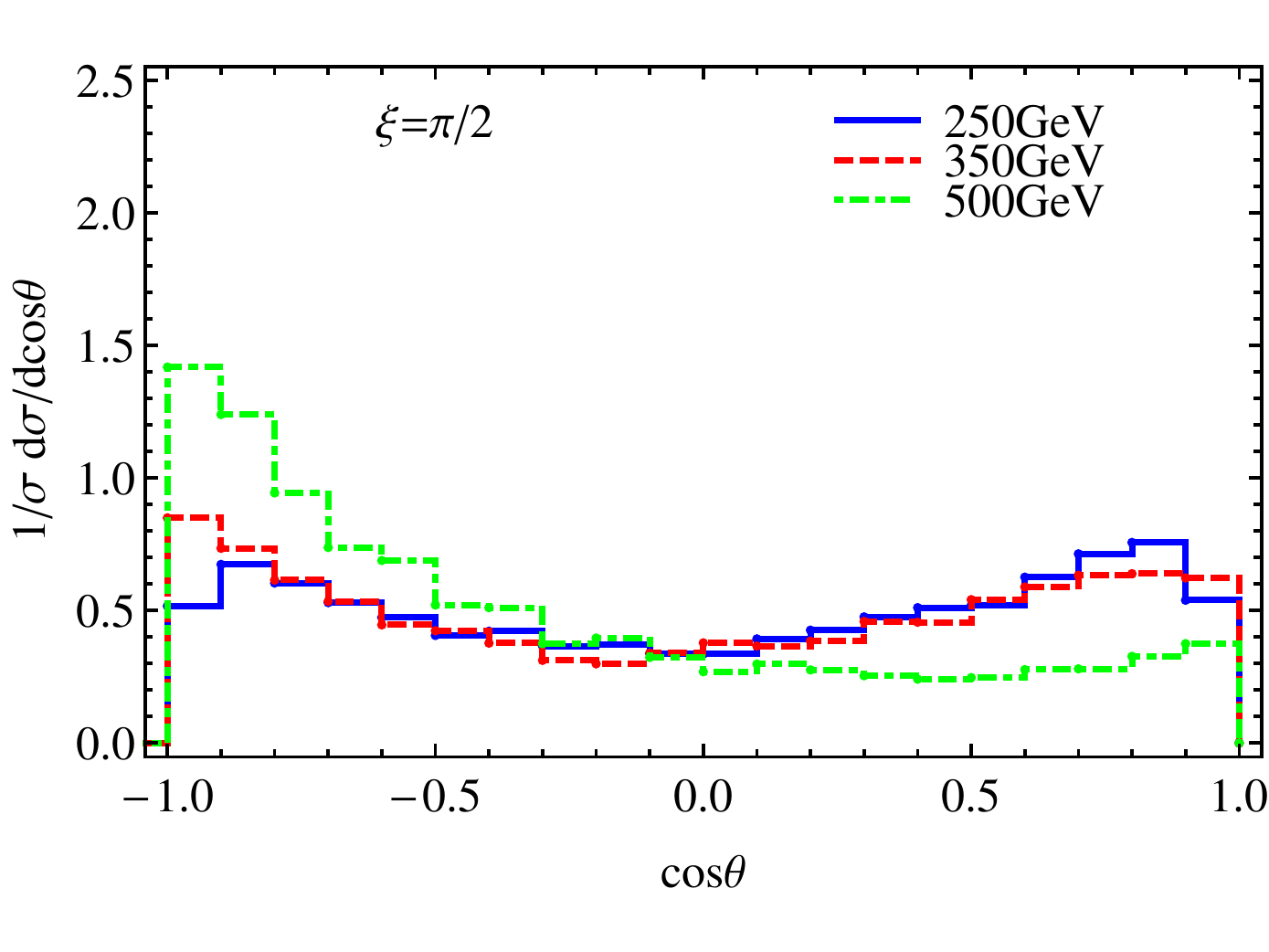}
\includegraphics[width=0.4\textwidth]{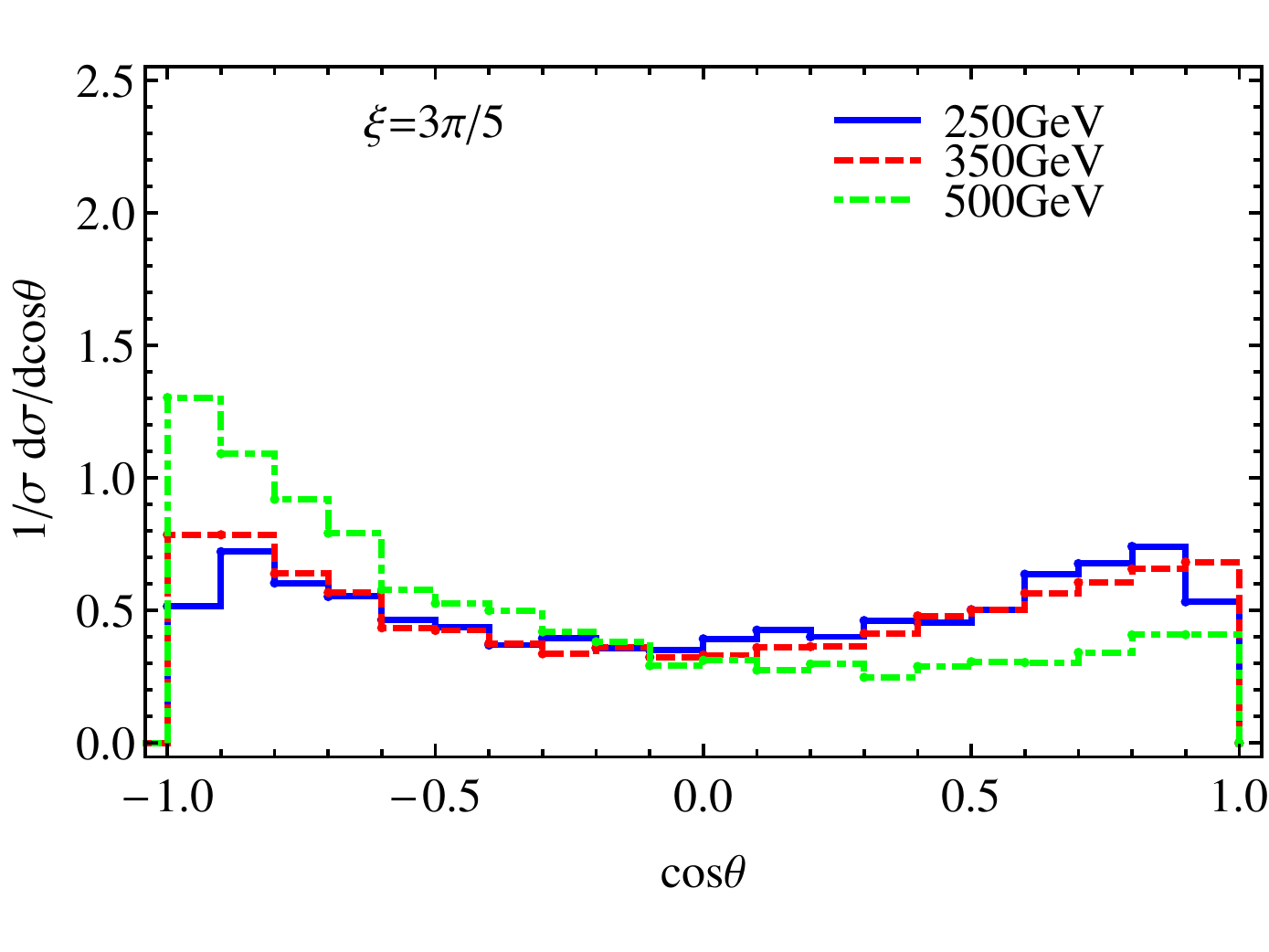}\\
   \caption{Angular distributions of the signal process ($\xi=0$, $\pi/4$, $\pi/2$, $3\pi/5$) that have passed the event selection cuts.}
 \label{fig:xdsex-mc}
   \end{figure}   
 \begin{figure}
  \centering
 \includegraphics[width=0.4\textwidth]{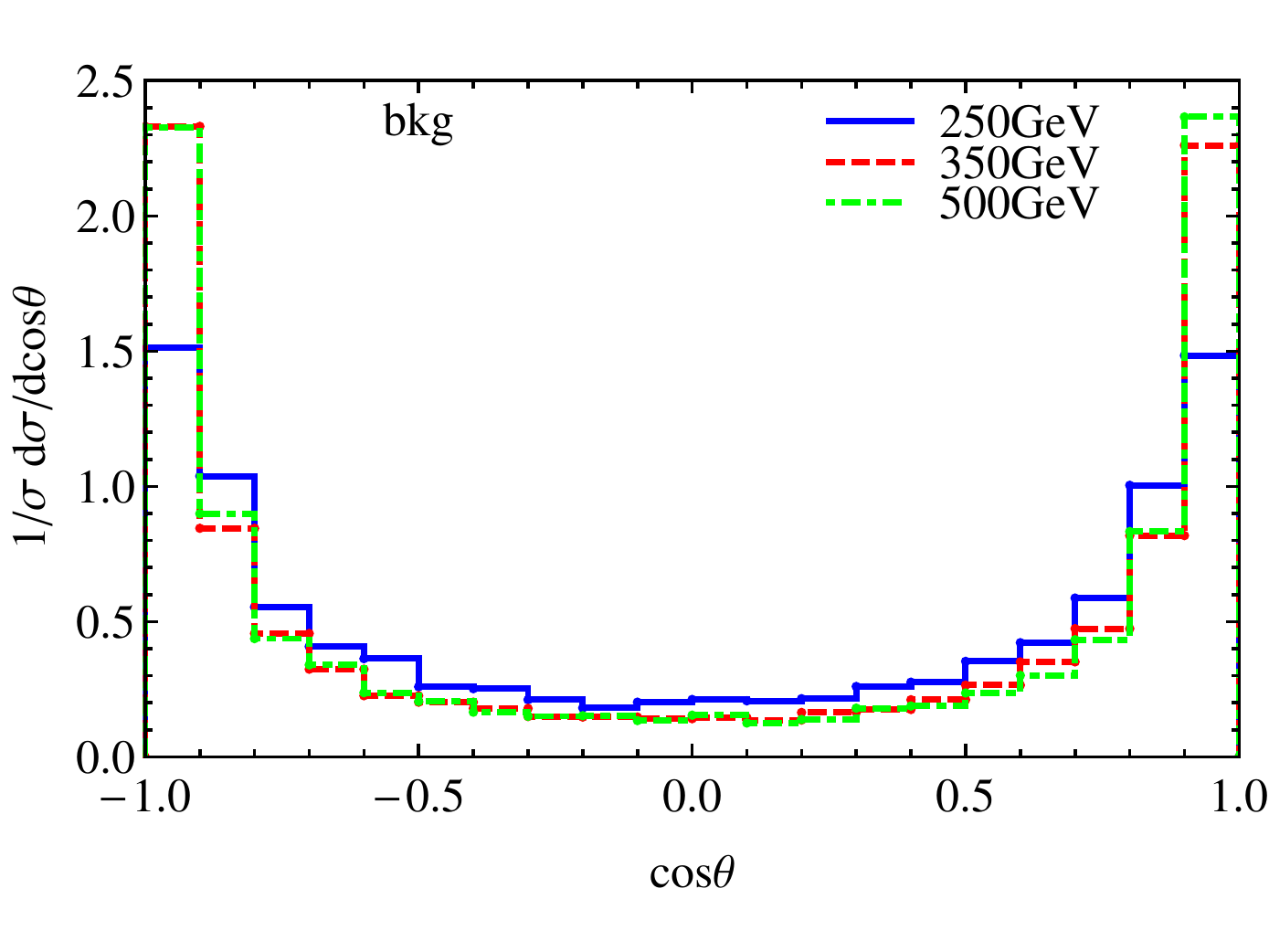}
   \caption{Angular distributions of the SM background (bkg) that has passed the event selection cuts. The distribution is symmetric.}
 \label{fig:xdsex-mc2}
   \end{figure}
    
\FIG{fig:xdsex-mc} shows the angular distributions of the signal processes $\eebbah$ with the CP phases $\xi=0,\ \pi/4,\ \pi/2,\ 3\pi/5$. The $p_T^\gamma$ and $\eta^\gamma$ cuts in \EQ{selectcut1} will constrain the maximum value of $\theta$ and hence reduce the heights of the peak in the forward/backward region, while the other cuts in~\EQS{selectcut1}~\eqref{selectcut2} are expected to have less impact on the distributions. The forward-backward asymmetries of the signal with different c.m. energies at a given value of $\xi$ are displayed in Table~\ref{tab:afb}. We find that the change after the cuts are applied is substantial at $\sqrt{s}=350~\rm{GeV}$, including the signs. This is because the $A_{FB}$ at $\sqrt{s}=350\GeV$ is very small and the sign is sensitive to the difference of events in the forward and backward regions. To illustrate the impact of the cuts on the  $A_{FB}$, it is helpful to write $A_{FB}={\Delta N_s}/{N_s}$ as in \EQ{afbgeneral}, where $\Delta N_s$ is the difference of the numbers of events in the forward and backward regions, and $N_s$ is the total number of events. The angular distributions at $\sqrt{s}=250~\rm{GeV}$ are nearly symmetric in \FIG{fig:dxsex}, so the cuts are expected to have more impact on $N_s$ rather than on $\Delta N_s$, and $A_{FB}$ will become larger in magnitude after the cuts are applied. On the other hand, the angular distributions at $\sqrt{s}=500~\rm{GeV}$ are apparently asymmetric. So the cuts will have more impact on $\Delta N_s$ and the magnitudes of $A_{FB}$ are reduced. For $\sqrt{s}=350~\rm{GeV}$, the distributions are moderately asymmetric, and $\Delta N_s$ is very sensitive to the curve shapes, so that the $A_{FB}$ are greatly affected by the cuts. \FIG{fig:xdsex-mc2} shows the angular distributions of the SM background process $e^{+}e^{-}\rightarrow b\bar{b}\gamma$, which remain nearly symmetric in $\cos\theta$~\cite{Renard:1981es} with the cuts being applied.

\begin{figure}
  \centering
  \includegraphics[width=0.4\textwidth]{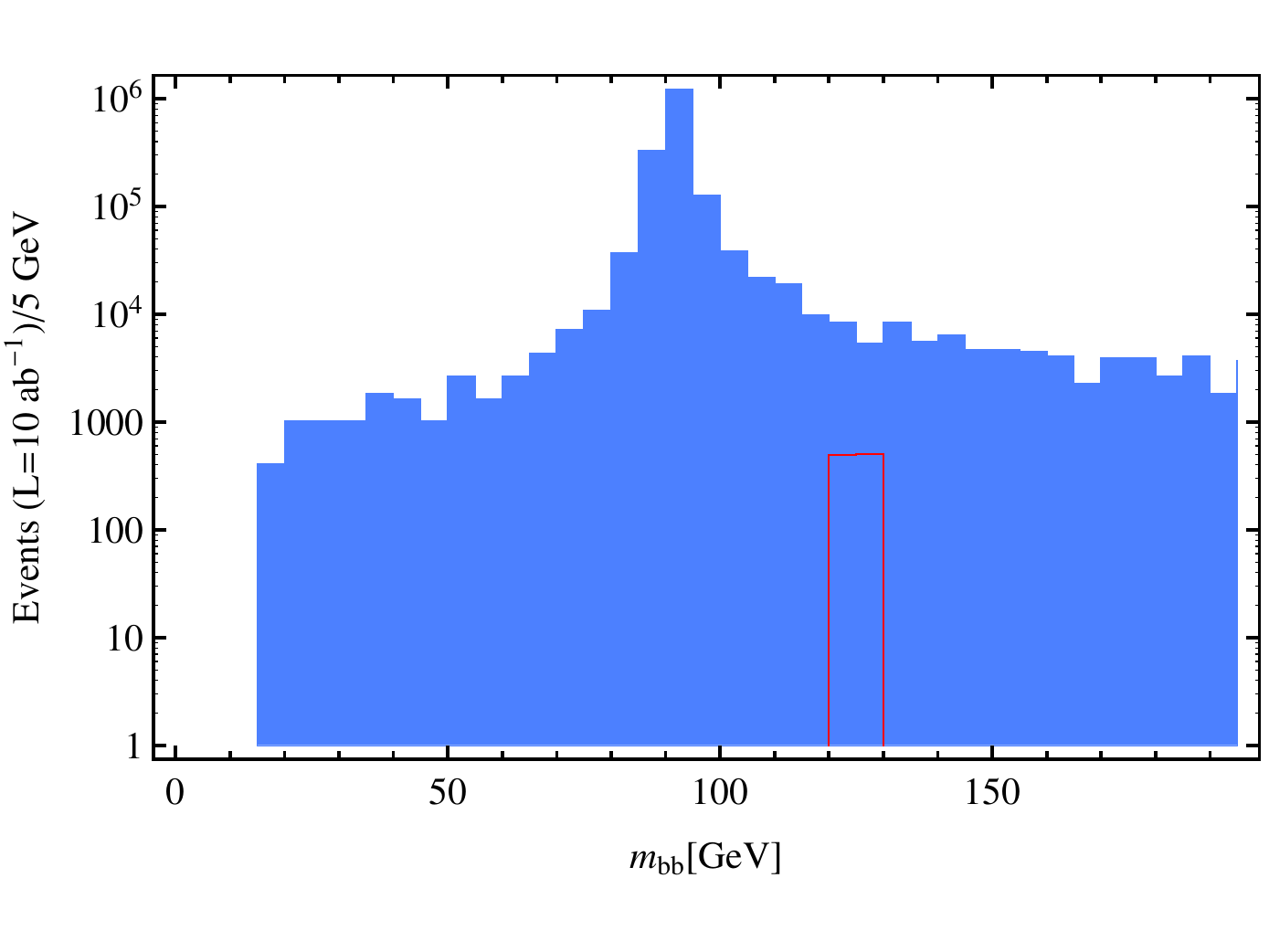}
 \includegraphics[width=0.4\textwidth]{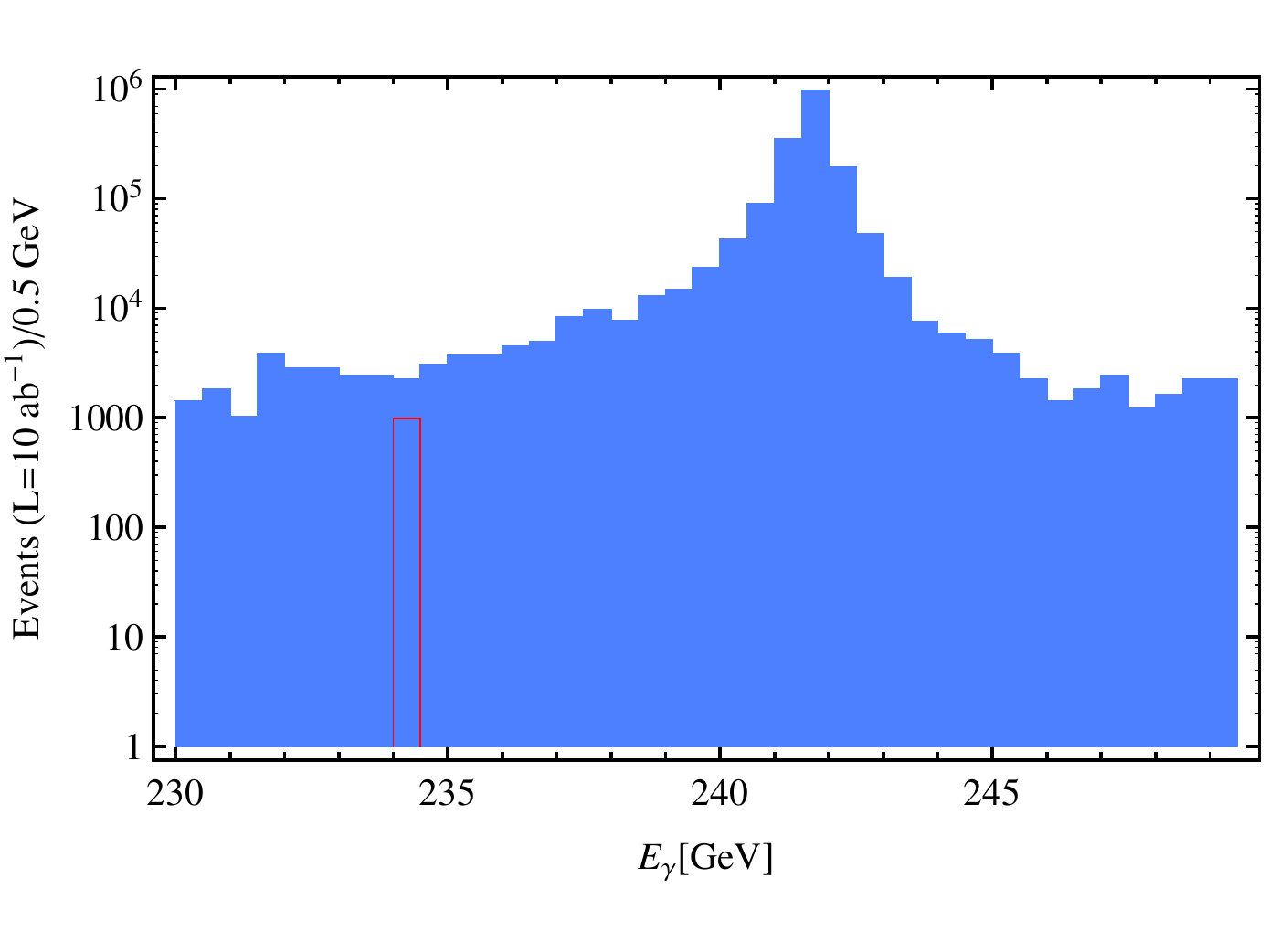}
   \caption{Left: the invariant mass distributions of the b jets. Right: the  distributions of the photon's recoil energy. The signal is in red, while the background is in blue. The events are collected for the integrated luminosity $\mathcal{L}=10~\text{ab}^{-1}$ and the c.m. energy $\sqrt{s}=500\GeV$.}
 \label{fig:photonenergy}
   \end{figure}

In Table~\ref{tab:s350} and Table~\ref{tab:s500}, we show the cutflows of the cross sections of the signals and background (bkg) at $\sqrt{s}=350\GeV,500\GeV$ and the significances corresponding to the integrated luminosity $\mathcal{L}=1~\rm{ab}^{-1}$, $3~\rm{ab}^{-1}$, $10~\rm{ab}^{-1}$. If there exists CP violating $h\bar{t}t$ interaction, the cross section at $\sqrt{s}=350\GeV$ is strongly enhanced. On the other hand, the magnitude of $A_{FB}$ at $\sqrt{s}=500\GeV$ is maximal which will be appropriate for the achievement of a larger statistical significance of the asymmetry. The foremost SM bkg is from the process $e^+e^-\to Z\gamma\to b\bar{b}\gamma$ and the behavior of its cross section is $\sim 1/s$~\cite{Renard:1981es}.
From the tables, we see that the bkg cross section is larger at $\sqrt{s}=350\GeV$ than at $\sqrt{s}=500\GeV$ before the photon energy resolution is required.
\FIG{fig:photonenergy} shows the signal and bkg distributions of the invariant mass of $b\bar{b}$ and the photon's recoil energy following the basic selection cuts (a) in \EQ{selectcut1}. So the mass window cut $\sim 15~\rm{GeV}$ and the requirement of the photon energy resolution are efficient to improve the signal significance. We have chosen the planed resolution $\Delta E_{\gamma}/E_{\gamma}\sim 1.7\%$~\cite{Behnke:2013lya} and the optimistic resolution $\Delta E_{\gamma}/E_{\gamma}\sim 0.5\%$~\cite{Cao:2015fra} in our analysis, and the remaining cross sections are displayed in Table~\ref{tab:s350} and Table~\ref{tab:s500}. We can find that the photon energy resolution has more impact on the bkg at $\sqrt{s}=350\GeV$ than at $\sqrt{s}=500\GeV$, due to the fact that $E_{\gamma}$ at $\sqrt{s}=350\GeV$ is smaller.
For the planned photon energy resolution, the signal significance $\mathcal{S}/\sqrt{\mathcal{B}}$ can reach $3\sigma$ with the integrated luminosity $\mathcal{L}=3~\rm{ab}^{-1}$ and about $5\sigma$ with $\mathcal{L}=10~\rm{ab}^{-1}$ at $\sqrt{s}=350\GeV$, and only $3\sigma$ with $\mathcal{L}=10~\rm{ab}^{-1}$ at $\sqrt{s}=500\GeV$. For the optimistic photon energy resolution, the significance increases to more than $3\sigma$ with $\mathcal{L}=1~\rm{ab}^{-1}$ and $5\sigma$ with $\mathcal{L}=3~\rm{ab}^{-1}$ at $\sqrt{s}=350\GeV$, and more than $2.1\sigma$ with $\mathcal{L}=3~\rm{ab}^{-1}$ and $4\sigma$ with $\mathcal{L}=10~\rm{ab}^{-1}$ at $\sqrt{s}=500\GeV$.
   \begin{center}
\captionof{table}{The forward-backward asymmetries of the signal before and after the cuts in \EQS{selectcut1}~\eqref{selectcut2} for different $\xi$ and $\sqrt{s}$.}
\label{tab:afb}
\begin{tabular}{ c|c|c|c}
\hline
$A_{FB}$ before cuts applied&$\sqrt{s}=250~\rm{GeV}$&$\sqrt{s}=350~\rm{GeV}$&$\sqrt{s}=500\rm{GeV}$\\
\hline
$\xi=\pi/4$&0.0392&-0.0231&-0.5478\\
\hline
$\xi=\pi/2$&0.0417&-0.0018&-0.4575\\
\hline
$\xi=3\pi/5$&0.0365&0.0029&-0.3987\\
\hline
\hline
$A_{FB}$ after cuts applied&$\sqrt{s}=250~\rm{GeV}$&$\sqrt{s}=350~\rm{GeV}$&$\sqrt{s}=500\rm{GeV}$\\
\hline
$\xi=\pi/4$&0.0464&0.0058&-0.5090\\
\hline
$\xi=\pi/2$&0.0596&0.0136&-0.4312\\
\hline
$\xi=3\pi/5$&0.0460&-0.0072&-0.3616\\
\hline
\end{tabular}
\end{center}
\begin{center}
\captionof{table}{Cutflow of the cross sections (in $\text{fb}$) of the signals with $\xi=0,\pi/4,\pi/2,3\pi/5$ and the background at $\sqrt{s}=350\GeV$ with the integrated luminosity $\mathcal{L}=1~\text{ab}^{-1}$, $3~\text{ab}^{-1}$ and $10~\text{ab}^{-1}$. The SM value $BR(h\to b\bar{b})=57.7\%$ and the b-tagging efficiency and the mis-tag probabilities in \EQ{efficiencies} have been included.}
\label{tab:s350}
\begin{tabular}{ c|cccc|c }
\hline
\multirow{2}{*}{$\sqrt{s}=350\GeV$}& \multicolumn{4}{|c|}{signals ($\mathcal{S}$)} & \multirow{2}{*}{bkg ($\mathcal{B}$)} \\
& $\xi=0~$ & $\xi=\pi/4~$ & $\xi=\pi/2~$ & $\xi=3\pi/5$ \\
\hline
cuts (a) in \EQ{selectcut1} & 0.017 &0.053 & 0.121&0.144 &477.55\\
\hline
$|m_{b\bar{b}}-m_{h}|\leq 15\GeV$ & 0.017 &0.053 & 0.121&0.144 &14.70\\
\hline
$\Delta E_{\gamma}< E_{\gamma}\times 1.7\%$ & 0.017 &0.053 & 0.121&0.144 &6.65\\
\hline
$\Delta E_{\gamma}< E_{\gamma}\times 0.5\%$ & 0.017 &0.053 & 0.121&0.144 &1.93\\
\hline
\hline
$\mathcal{S}/\mathcal{B}$ & 0.003 &0.008 &0.018 &0.022 &\multirow{4}{*}{6.65}\\
\hhline{-----~}
$\mathcal{S}/\sqrt{\mathcal{B}}$ with $1~\text{ab}^{-1}$&0.212&0.645 &1.488 &1.762& \\
\hhline{-----~}
$\mathcal{S}/\sqrt{\mathcal{B}}$ with $3~\text{ab}^{-1}$&0.367 &1.118 &2.577 &3.052 &\\
\hhline{-----~}
$\mathcal{S}/\sqrt{\mathcal{B}}$ with $10~\text{ab}^{-1}$&0.669&2.041&4.705&5.573&\\
\hline
\hline
$\mathcal{S}/\mathcal{B}$ & 0.009 &0.027 &0.063 &0.074 &\multirow{4}{*}{1.93}\\
\hhline{-----~}
$\mathcal{S}/\sqrt{\mathcal{B}}$ with $1~\text{ab}^{-1}$&0.393&1.197 &2.760 &3.269 &\\
\hhline{-----~}
$\mathcal{S}/\sqrt{\mathcal{B}}$ with $3~\text{ab}^{-1}$&0.680 &2.073 &4.781 &5.662 &\\
\hhline{-----~}
$\mathcal{S}/\sqrt{\mathcal{B}}$ with $10~\text{ab}^{-1}$&1.241&3.785&8.728&10.338&\\
\hline
\end{tabular}
\end{center}

\begin{center}
\captionof{table}{Cutflow of the cross sections (in $\text{fb}$) of the signals with $\xi=0,\pi/4,\pi/2,3\pi/5$ and the background at $\sqrt{s}=500\GeV$ with the integrated luminosity $\mathcal{L}=1~\text{ab}^{-1}$, $3~\text{ab}^{-1}$ and $10~\text{ab}^{-1}$. The SM value $BR(h\to b\bar{b})=57.7\%$ and the b-tagging efficiency and the mis-tag probabilities in \EQ{efficiencies} have been included.}
\label{tab:s500}
\begin{tabular}{ c|cccc|c }
\hline
\multirow{2}{*}{$\sqrt{s}=500\GeV$}& \multicolumn{4}{|c|}{signals ($\mathcal{S}$)} & \multirow{2}{*}{bkg ($\mathcal{B}$)} \\
& $\xi=0~$ & $\xi=\pi/4~$ & $\xi=\pi/2~$ & $\xi=3\pi/5$ \\
\hline
cuts (a) in \EQ{selectcut1} & 0.026 &0.041 & 0.067&0.073 &225.53\\
\hline
$|m_{b\bar{b}}-m_{h}|\leq 15\GeV$ & 0.026 &0.041 & 0.067&0.073 &6.12\\
\hline
$\Delta E_{\gamma}< E_{\gamma}\times 1.5\%$ & 0.026 &0.041 & 0.067&0.073 &5.86\\
\hline
$\Delta E_{\gamma}< E_{\gamma}\times 0.5\%$ & 0.026 &0.041 & 0.067&0.073 &2.90\\
\hline
\hline
$\mathcal{S}/\mathcal{B}$ &0.004 &0.007 &0.011 &0.012 &\multirow{4}{*} {5.86}\\
\hhline{-----~}
$\mathcal{S}/\sqrt{\mathcal{B}}$ with $1~\text{ab}^{-1}$&0.337 &0.540 &0.875 &0.951 &\\
\hhline{-----~}
$\mathcal{S}/\sqrt{\mathcal{B}}$ with $3~\text{ab}^{-1}$&0.584 &0.936 &1.515 &1.647 &\\
\hhline{-----~}
$\mathcal{S}/\sqrt{\mathcal{B}}$ with $10~\text{ab}^{-1}$&1.067 &1.709 &2.766 &3.008 &\\
\hline
\hline
$\mathcal{S}/\mathcal{B}$ &0.009 &0.014 &0.023 &0.025 &\multirow{4}{*}{2.90}\\
\hhline{-----~}
$\mathcal{S}/\sqrt{\mathcal{B}}$ with $1~\text{ab}^{-1}$&0.480 &0.769 &1.244 &1.353 &\\
\hhline{-----~}
$\mathcal{S}/\sqrt{\mathcal{B}}$ with $3~\text{ab}^{-1}$&0.831 &1.332 &2.155 &2.344 &\\
\hhline{-----~}
$\mathcal{S}/\sqrt{\mathcal{B}}$ with $10~\text{ab}^{-1}$&1.517 &2.431 &3.935 &4.279 &\\
\hline
\end{tabular}
\end{center}

In order to estimate whether the $\AFB$ can be measured at the future high luminosity $e^{+}e^{-}$ colliders, we calculate the significance of the expected asymmetry with which a particular CP phase $\xi$ would manifest~\cite{Godbole:2007cn}
\beq
\label{significance}
S=|\AFB |\frac{N_{s}}{\sqrt{N_{s}+N_{b}}},
\eeq
where $N_{s}$, $N_{b}$ are the number of signal and background events, respectively, and $\AFB$ is the theoretical asymmetry given in~\EQ{afbexact}. The significances at $\sqrt{s}=500\GeV$ are larger than those at $\sqrt{s}=350\GeV$. For the c.m. energy  $\sqrt{s}=500\GeV$ and the integrated luminosity $\mathcal{L}=10~\text{ab}^{-1}$, the significances are given by $S=1.23\sigma,\ 1.68\sigma,\ 1.57\sigma$ for $\xi=\pi/4,\ \pi/2,\ 3\pi/5$, respectively.

\section{Conclusions and Discussions}
In this paper, we have investigated the effects of the CP-violating $h\bar t t$ coupling in the process $\eeha$. Our numerical results show  that the cross section can increase significantly for the allowed CP phase $\xi$ and the c.m. energy. For $\sqrt{s}=350\GeV$ and $\xi =3\pi/5$, the cross section is about 10 times of that
in the SM. A preliminary simulation for signal $\eebbah$ and its backgrounds has been carried out. 
For $\sqrt{s}=350\GeV$, we can observe the signal at about $5\sigma$ with the integrated luminosity $\mathcal{L}=3~\text{ab}^{-1}$ with $\xi\in[\pi/2,3\pi/5]$. For $\sqrt{s}=500\GeV$, we can observe the signal at more than $2.1\sigma$ with $\mathcal{L}=3~\text{ab}^{-1}$  with $\xi\in[\pi/2,3\pi/5]$. For $\mathcal{L}=10~\text{ab}^{-1}$, the significance can be greater than $5\sigma$ for $\sqrt{s}=350\GeV$ and than $4\sigma$ for $ 500\GeV$ with $\xi\in[\pi/2,3\pi/5]$.

Compared with the $\AFB$ in the Higgs decay $h\rightarrow l^{+}l^{-}\gamma$, the $\AFB$ can be greatly enhanced in the production process. $\AFB$ can reach -0.55 for $\xi=\pi/4$ and $\sqrt{s}=500\GeV$. Due to the large backgrounds, the significance of the expected $\AFB$ can be observed at only $1.68\sigma$ with $\mathcal{L}=10~\text{ab}^{-1}$ and $\sqrt{s}=500\GeV$. We should emphasize that it is essential to trigger the single photon in the final state to separate the bottom jets arising from scalar or vector bosons,
in order to isolate the signal from backgrounds, especially for high c.m. energy.
\begin{acknowledgments}
We would like to thank Qing-Hong Cao, Yin-nan Mao, Hong-Yu Ren, Bin Yan, Chen Zhang and Dong-Ming Zhang for useful discussions. GL would also thank Shi Ang Li and Yong Chuan Zhan for the help with loop calculations. This work was supported in part by the Natural Science Foundation of
China (Grants No. 11135003 and No. 11375014).
\end{acknowledgments}

\appendix
\section{Analytical expression of scalar functions }
\label{Appdendix:analyticalexpression}
In this appendix, we give the analytical expressions of two-point and three-point scalar functions $B_{0}$ and $C_{0}$.
 The scalar functions $B_{0}$ and $C_{0}$ are defined as \cite{Denner:1991kt}
 \bea
  B_{0}(p_1^2,m_1^{2},m_2^{2})
  =\frac{1}{i\pi^{2}}(2\pi\mu)^{4-D}\int d^{D}k\frac{1}{[k^{2}-m_1^{2}][(k+p_1)^{2}-m_2^{2}]},\nn\\
  C_{0}(p_1^2,p_2^2,p_3^2,m_1^{2},m_2^{2},m_3^{2})
  =\frac{1}{i\pi^{2}}(2\pi\mu)^{4-D}\int d^{D}k\frac{1}{[k^{2}-m_1^{2}][(k+p_1)^{2}-m_2^{2}][(k+p_1+p_2)^{2}-m_3^{2}]},
 \eea
 and the analytical expressions are \cite{'tHooft:1978xw, Fortes:2014dia}
 \bea
  B_{0}(s,m^{2},m^{2})&=&\Delta_{\epsilon}-\int_{0}^{1}dx\log\frac{-x(1-x)s+xs+(1-x)m^{2}}{\mu^{2}}\nn\\
                              &=&\Delta_{\epsilon}-\log\frac{m^{2}}{\mu^{2}}+2-2g(\lambda),\nn\\
  B_{0}(m_{h}^{2},m^{2},m^{2})&=&\Delta_{\epsilon}-\int_{0}^{1}dx\log\frac{-x(1-x)m_{h}^{2}+x m_{h}^{2}+(1-x)m^{2}}{\mu^{2}}\nn\\
                              &=&\Delta_{\epsilon}-\log\frac{m^{2}}{\mu^{2}}+2-2g(\tau),\nn\\
  C_{0}(0,0,m_{h}^{2},m^{2},m^{2},m^{2})&=&-\int_{0}^{1}dx\int_{0}^{1-x}dy\frac{1}{m_{h}^{2}(x^{2}-xy-x+y)+m^{2}-i\varepsilon}\nn\\
                                        &=&-\frac{2}{m_{h}^{2}}f(\tau),\nn\\
  C_{0}(0,0,s,m^{2},m^{2},m^{2})&=&-\int_{0}^{1}dx\int_{0}^{1-x}dy\frac{1}{s(x^{2}-xy-x+y)+m^{2}-i\varepsilon}\nn\\
                                        &=&-\frac{2}{s}f(\lambda),\nn\\
  C_{0}(s,0,m_{h}^{2},m^{2},m^{2},m^{2})&=&-\int_{0}^{1}dx\int_{0}^{1-x}dy\frac{1}{(s-m_{h}^{2})xy+m_{h}^{2}x(x-1)+m^{2}-i\varepsilon}\nn\\
                                        &=&\frac{1}{4m^{2}}\frac{-2\tau\lambda}{\lambda-\tau}(f(\tau)-f(\lambda))
 \eea
 where $\Delta_{\epsilon}=\frac{2}{\epsilon}-\gamma_{E}+\log4\pi,\epsilon=4-D$, $\gamma_{E}=0.5772\cdots$ is the Euler constant, $\tau=\frac{4m^{2}}{m_{h}^{2}},\lambda=\frac{4m^{2}}{s}$ and the functions \cite{Djouadi:2005gi}
\bea
\label{eq:f_x}
f(\tau)&=&
\begin{cases}
\arcsin^{2}\sqrt{\tau-1}, &  \tau \geq 1,\\
-\frac{1}{4}\sqrt{1-\tau}(\log\frac{\eta_{+}}{\eta_{-}}-i\pi)^{2} &  0<\tau < 1,\\
\end{cases}
\eea
\bea
\label{eq:g_x}
g(\tau)&=&
\begin{cases}
\sqrt{\tau-1}\arcsin^{2}\sqrt{\tau^{-1}}, &  \tau \geq 1,\\
\frac{1}{2}\sqrt{1-\tau}(\log\frac{\eta_{+}}{\eta_{-}}-i\pi) &  0<\tau < 1,\\
\end{cases}
\eea
 with $\eta_{\pm}=1\pm\sqrt{1-\tau}$.
\section{Loop functions and form factors}
\label{loopfunctions}
In this appendix, we will give the detailed s-channel $W$-boson loop functions and t-channel and box diagram form factors of the process $e^+e^-\to h\gamma$, computed in \cite{Djouadi:1996ws}.
\subsection{s-channel diagram contributions}
$W$ boson loop functions
\bea
F_1^{\gamma,W}&=&(\frac{m_h^2}{m_W^2}+6)(4C_{24}-B_{13})+m_h^2(-7C_0+C_{11}+C_{12})\nn\\
&+&s(5C_0+C_{11}-C_{12})-B_{12}+B_{23},\\
F_3^{\gamma,W}&=&4(\frac{m_h^2}{m_W^2}+6)(C_{12}+C_{23})+16C_{0},\\
F_1^{Z,W}&=&\frac{1}{2}\frac{m_h^2}{m_W^2}(1-2c_W^2)(4C_{24}-B_{13})+m_h^2c_W^2(7C_0-C_{11}-C_{12}-2C_{23})\nn\\
&+&m_h^2(-C_0+C_{11}-C_{23})+2m_W^2(c_W^2-1)C_0\nn\\
&-&c_W^2[s(5C_0+C_{11}-C_{12}+2C_{21}-2C_{23})+32C_{24}-B_{12}-6B_{13}-B_{23}]\nn\\
&+&\frac{1}{2}+c_W^2+s(C_0-C_{11}-C_{21}+C_{23})-B_{12}+B_{23},\\
F_3^{Z,W}&=&2[\frac{m_h^2}{m_W^2}(1-2c_W^2)+2(1-6c_W^2)](C_{12}+C_{23})+4(1-4c_W^2)C_0,
\eea
where the two-point and three-point functions are
\bea
B_{12}&=&B_0(s,m_W^2,m_W^2),\ B_{13}=B_0(m_h^2,m_W^2,m_W^2),\nn\\
B_{23}&=&B_0(0,m_W^2,m_W^2),\ C_{0;ij}=C_{0;ij}(s,0,m_h^2,m_W^2,m_W^2,m_W^2).
\eea
\subsection{t-channel diagram contributions}
\begin{align*}
-C_i^{e\pm}&=\frac{e^4}{s_W^3}[\frac{m_W}{2}A_i^{W\pm}+\frac{m_Z}{4c_W^3}A_i^{Z\pm}]+\rm{crossed}~(i=1,2,3),\\
C_6^{e\pm}&=0,
\end{align*}
where
\bea
A_3^{W+}&=&C_{12}(m_e^2,t,m_h^2,m_W^2,0,m_W^2),\ A_3^{W-}=0,\ A_1^{W\pm}=A_2^{W\pm}=0,\nn\\
A_3^{Z\pm}&=&(z^{\pm})^2C_{12}(m_e^2,t,m_h^2,m_Z^2,m_e^2,m_Z^2),\ A_1^{Z\pm}=A_2^{Z\pm}=0
\eea
and the crossed terms are obtained by substituting $t\to u$.
\subsection{box diagram contributions}
\bea
 \label{box1}
C_1^{\rm{box}\pm}&=&-\frac{e^4m_W}{4s_W^3}[B_1^{W\pm}+\text{crossed}(B_2^{W\pm})]+\frac{e^4m_Z}{4s_W^3c_W^3}B_1^{Z\pm},\\
 \label{box2}
C_2^{\rm{box}\pm}&=&-\frac{e^4m_W}{4s_W^3}[B_2^{W\pm}+\text{crossed}(B_1^{W\pm})]+\frac{e^4m_Z}{4s_W^3c_W^3}B_2^{Z\pm},\\
-C_3^{\rm{box}\pm}&=&-\frac{e^4m_W}{4s_W^3}[B_3^{W\pm}+\text{crossed}(B_3^{W\pm})]+\frac{e^4m_Z}{4s_W^3c_W^3}B_3^{Z\pm},\\
C_6^{\rm{box}\pm}&=&0,
\eea
where\footnote{There are several typos in the above expressions in Ref.~\cite{Djouadi:1996ws} which have been corrected here.}
\bea
B_1^{W+}&=&4(D_0+D_{11}-D_{13}+D_{23}-D_{25}),\nn\\
B_2^{W+}&=&4(-D_{12}+D_{13}+D_{23}-D_{26}),\nn\\
B_3^{W+}&=&\frac{1}{2}[3s(-D_{12}+D_{13}+D_{25}+D_{26}-D_{23}-D_{24})\nn\\
&+&7t(-D_{23}+D_{26})+u(3D_{13}+7D_{25}-7D_{23})-20D_{27}],\nn\\
B_{1,2,3}^{W-}&=&0,
\eea
with the four-point functions
\beq
D_{0;ij}=D_{0;ij}(m_e^2,m_e^2,m_h^2,0,s,u,m_W^2,0,m_W^2,m_W^2),
\eeq
and 
\bea
B_1^{Z\pm}&=&2(z^{\pm})^2(-D_{11}+D_{12}+D_{22}-D_{24}),\nn\\
B_2^{Z\pm}&=&2(z^{\pm})^2(D_{22}-D_{26}),\nn\\
B_3^{Z\pm}&=&(z^{\pm})^2[s(D_{22}-D_{24}+D_{25}-D_{26})+2D_{27}]
\eea
with
\beq
D_{0;ij}=D_{0;ij}(m_e^2,m_h^2,m_e^2,0,t,u,m_e^2,m_Z^2,m_Z^2,m_e^2).
\eeq
The crossed terms are obtained by substituting $t\leftrightarrow u$. In fact, we find that
\begin{align}
\label{crossed}
B_2^{Z\pm}&=\text{crossed}(B_1^{Z\pm}), & C_2^{\rm{box}\pm}&=\text{crossed}(C_1^{\rm{box}\pm}),
\end{align}
and the combination $C_1^{\rm{box}\pm}+C_2^{\rm{box}\pm}$ is symmetric in $\cos\theta$.

    \begin{figure}
  \centering
 \includegraphics[width=0.3\textwidth]{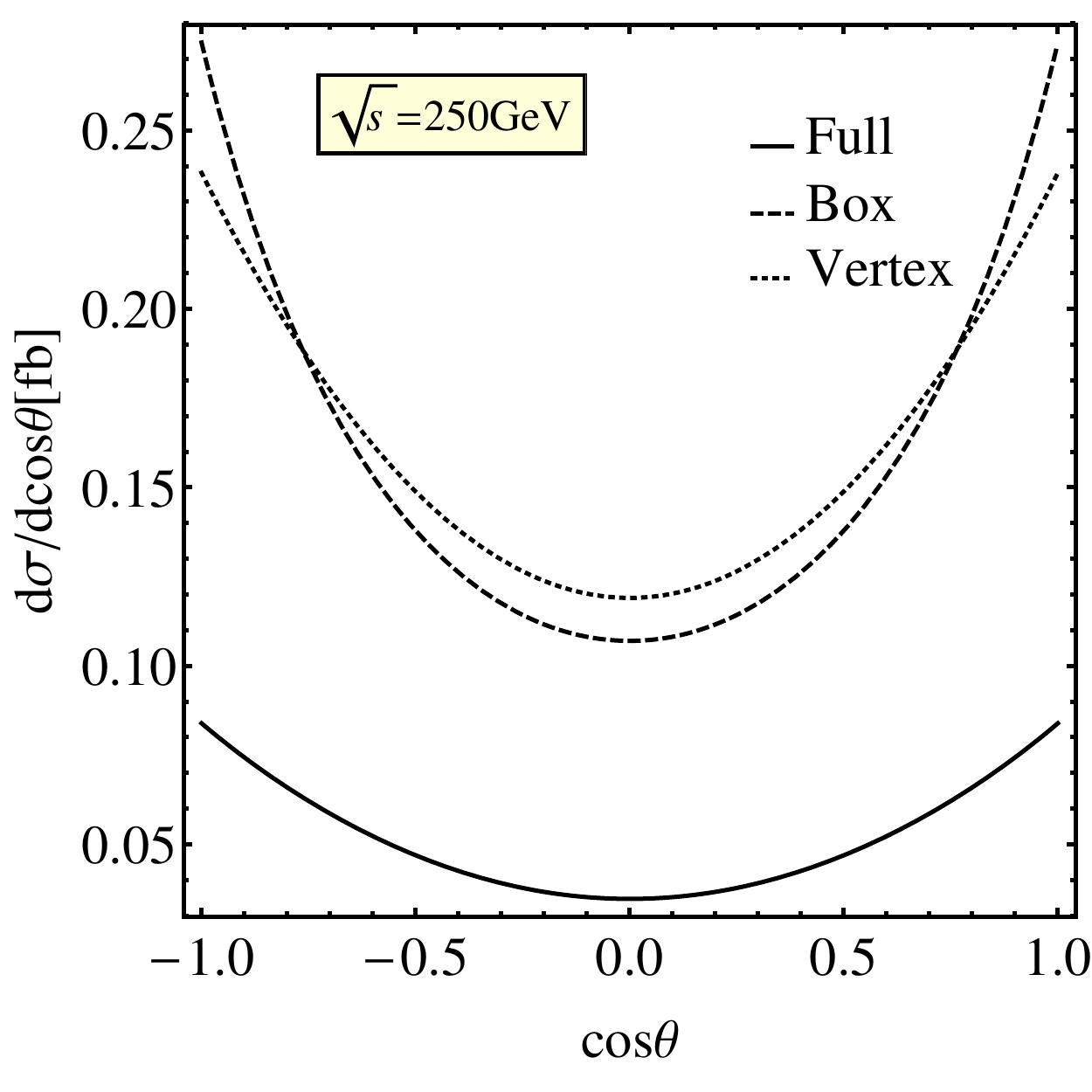}
  \includegraphics[width=0.3\textwidth]{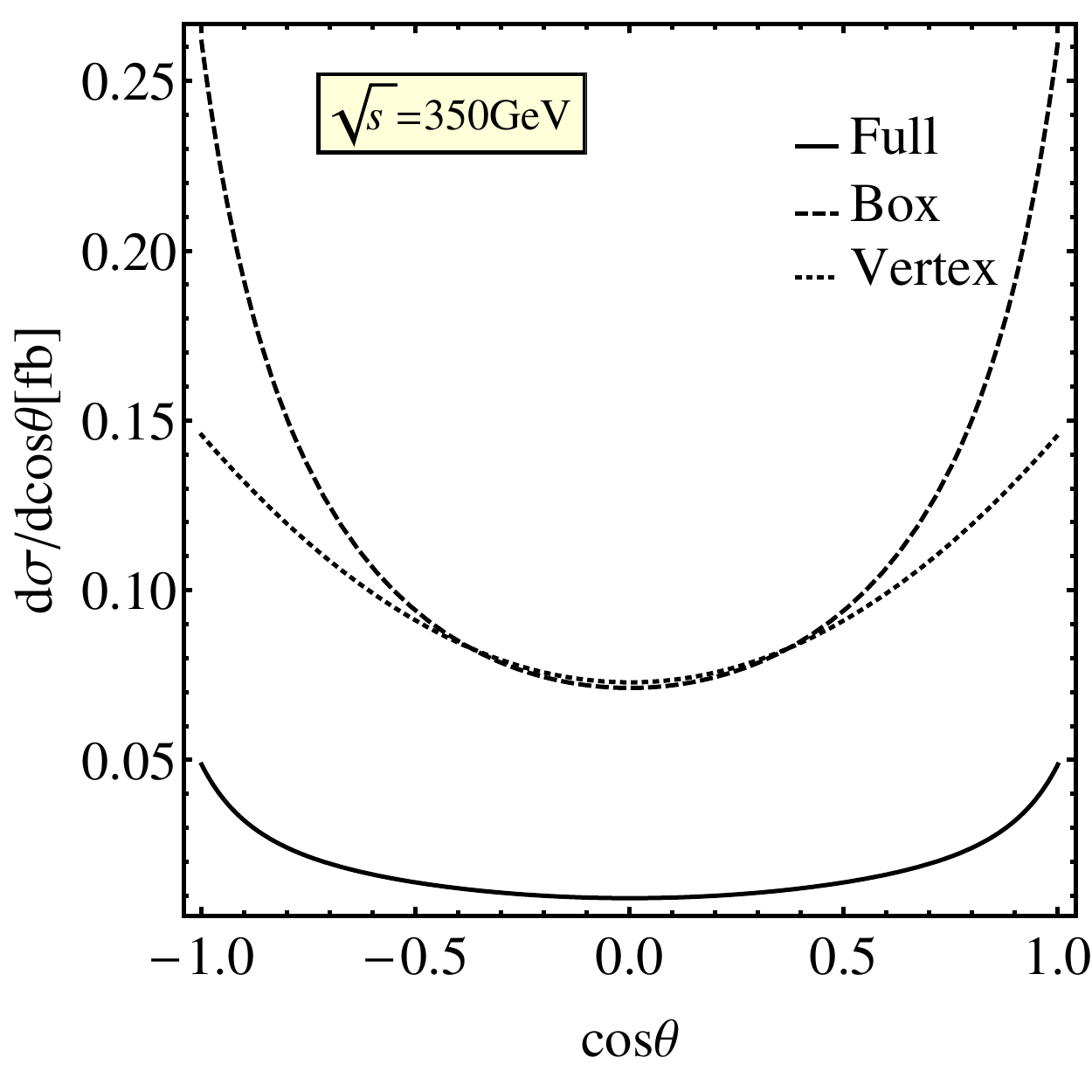}
  \includegraphics[width=0.3\textwidth]{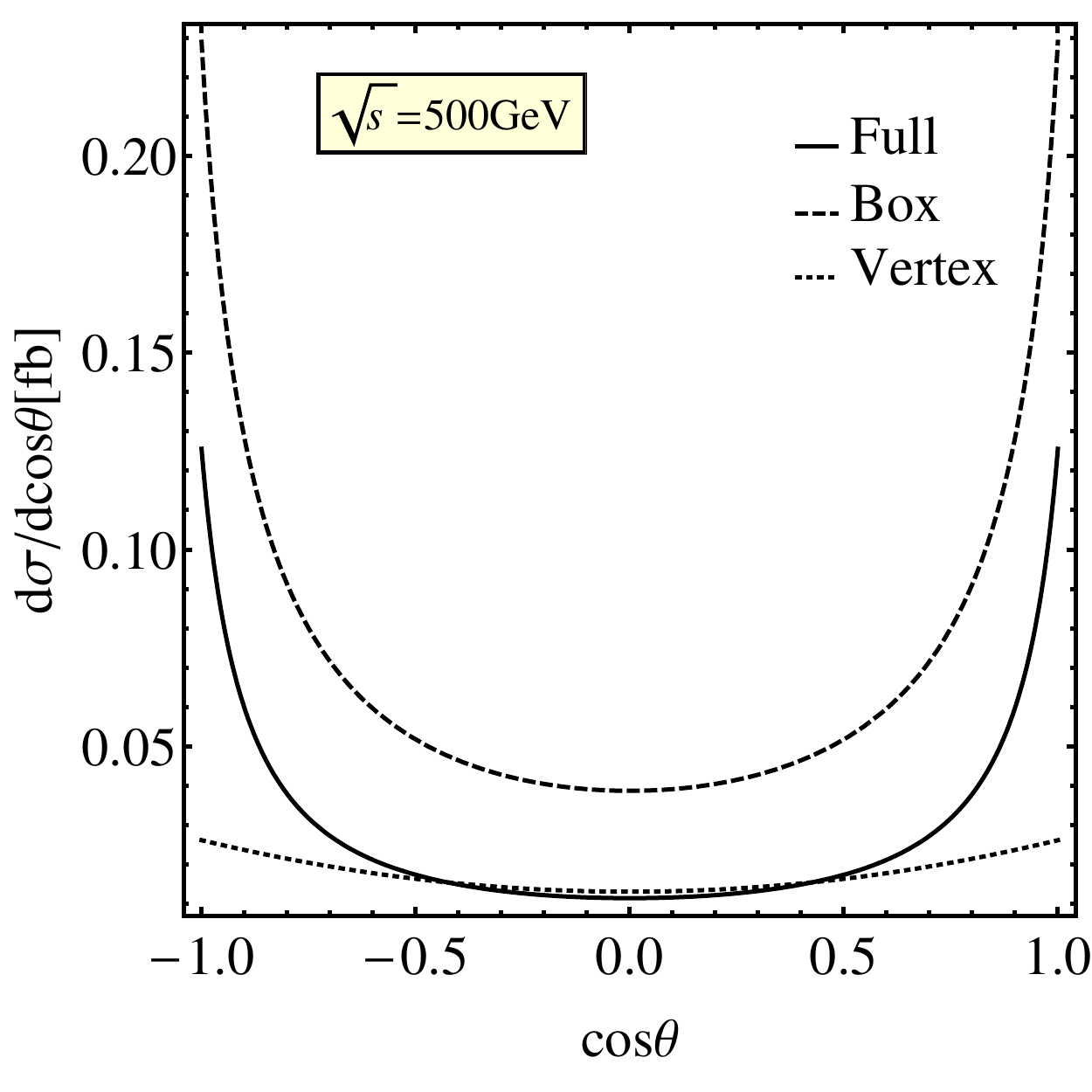}
   \caption{The differential cross sections $d\sigma/d\cos\theta$ at c.m. energies $\sqrt{s}=250\GeV$, $250\GeV$ and $500\GeV$ in the SM. The solid, dashed and dotted curves denote the full, box diagram and s-channel vertex diagram contributions, respectively.}
 \label{fig:plotbox}
   \end{figure}
In \FIG{fig:plotbox}, we show the SM ($\xi=0$) the differential cross sections $d\sigma/d\cos\theta$ at c.m. energies $\sqrt{s}=250\GeV$, $250\GeV$ and $500\GeV$. The solid, dashed and dotted curves denote the full, box diagram and s-channel vertex diagram contributions, respectively. We can easily find that the box diagram contributions interfere destructively with the s-channel vertex diagram contributions and dominate at higher c.m. energy in the SM. 

Finally, we have checked explicitly the following relations using the LoopTools package~\cite{Hahn:1998yk},
\bea
\label{gi1}
C_3^{\gamma-}(f)=C_3^{Z-}(f)=C_3^{\gamma+}(f)=C_3^{Z+}(f)=0,\nn\\ 
C_3^{\gamma-}(W)+C_3^{Z-}(W)=0,\nn\\
C_3^{e-}+C_3^{\rm{box}-}=0,\nn\\
C_3^{\gamma+}(W)+C_3^{Z+}(W)+C_3^{e+}+C_3^{\rm{box}+}=0,
\eea
which are the results of gauge invariance, and the form factors $C_3^{\gamma,Z\pm}(f)$ and $C_3^{\gamma,Z\pm}(W)$ denote the fermion-loop and $W$-loop contributions to $C_3^{\gamma,Z\pm}$, respectively. 
\section{Significance of the forward-backward asymmetry }
\label{Appendix:significance}
In this appendix, we will give the detailed derivation of the significance of the expected asymmetry in~\EQ{significance}, which has been discussed partly in in~\cite{Godbole:2007cn,BhupalDev:2007is,Godbole:2011hw,
Belyaev:2015xwa,Chen:2014ona}.We define the theoretical asymmetry $A_{\rm{FB}}$ and the measured asymmetry $A_{\rm{FB}}^{\rm{meas}}$ as
\bea
\label{afbgeneral}
A_{\rm{FB}}&=&\frac{\Delta N_s}{N_s},\\
A_{\rm{FB}}^{\rm{meas}}&=&\frac{\Delta N}{N_s+N_b},
\eea
where $\Delta N_s=N_s^F-N_s^B$ and $N_s^F$ and $N_s^B$ are the number of events in the forward and backward regions, respectively. $N_s= N_s^F+N_s^B$ is the total number of signal events, and $N_b$ is the number of the background events. $\Delta N\simeq\Delta N_s$ if the background does not contribute to the asymmetry or the contribution is much smaller. Write $N_s^F=F$ and $N_s^B=B$, we have
\beq
A_{\rm{FB}}=\frac{F-B}{F+B}.
\eeq
The statistical error of $A_{\rm{FB}}$ is expressed as the error propagation,
\beq
\sigma_{A}^2=(\frac{\partial A_{\rm{FB}}}{\partial F})^2\sigma_F^2+(\frac{\partial A_{\rm{FB}}}{\partial B})^2\sigma_B^2.
\eeq
For a Poison distribution, $\sigma_F^2=F$ and $\sigma_B^2=B$. Thus we obtain
\beq
\sigma_A^2=\frac{1-A_{\rm{FB}}^{2}}{N_s},
\eeq
which is approximate $\sigma_A=1/\sqrt{N_s}$ for a small $A_{\rm{FB}}$. Similarly, for the measured asymmetry $A_{\rm{FB}}^{\rm{meas}}$ the statistical error is 
\beq
\label{staterr}
(\sigma_A^{\rm{meas}})^2=\frac{1-(A_{\rm{FB}}^{\rm{meas}})^{2}}{N_s+N_b}\simeq \frac{1}{N_s+N_b}.
\eeq
The significance of the expected asymmetry $S$ is defined in~\cite{Godbole:2007cn}
\beq
S=\frac{|A_{\rm{FB}}^{\rm{meas}}|}{\sigma_A^{\rm{meas}}},
\eeq
or in Refs.~\cite{Belyaev:2015xwa,Accomando:2015cfa}
\beq
S=\frac{\Delta N}{\sqrt{N_s+N_b}}.
\eeq
In both cases,
\beq
S=|A_{\rm{FB}}^{\rm{meas}}|\sqrt{N_s+N_b}
 \simeq |A_{\rm{FB}}|\frac{N_s}{\sqrt{N_s+N_b}}.
\eeq

\bibliographystyle{apsrev}
\bibliography{reference}

\end{document}